\documentclass[aps,prb,twocolumn,groupedaddress,eqsecnum,draft,showpacs,intlimits,amsmath,amssymb,floats]{revtex4}
\usepackage{bm}
\usepackage[final]{graphicx}
 \DeclareMathOperator{\Img}{Im}
\DeclareMathOperator{\Real}{Re} \DeclareMathOperator{\sgn}{sgn}
\DeclareMathOperator{\erfc}{erfc} \righthyphenmin=3

\begin{document}
\title{Electronic Raman scattering in correlated materials: exact treatment
of nonresonant, mixed, and resonant scattering with dynamical mean field
theory}
\date{\today}
\author{A. M. Shvaika}
\email{ashv@icmp.lviv.ua} \homepage{http://ph.icmp.lviv.ua/~ashv/}
\affiliation{Institute for Condensed Matter Physics of the National
Academy of Sciences of Ukraine, 1 Svientsitskii Street, 79011 Lviv,
Ukraine}
\author{O. Vorobyov}
\email{vorobyov@icmp.lviv.ua} \homepage{http://ph.icmp.lviv.ua/~vorobyov/}
\affiliation{Institute for Condensed Matter Physics of the National
Academy of Sciences of Ukraine, 1 Svientsitskii Street, 79011 Lviv,
Ukraine}
\author{J. K. Freericks}
\email{freericks@physics.georgetown.edu}
\homepage{http://www.physics.georgetown.edu/~jkf/} \affiliation{Department
of Physics, Georgetown University, Washington, DC 20057}
\author{T. P. Devereaux}
\email{tpd@lorax.uwaterloo.ca}
\homepage{http://www.sciborg.uwaterloo.ca/~tpd/} \affiliation{Department
of Physics, University of Waterloo, Canada, Ontario N2L 3GI}

\begin{abstract}
We solve for the electronic Raman scattering response functions on an
infinite-dimensional hypercubic lattice employing dynamical mean field
theory.  This contribution extends previous work on the nonresonant
response to include the mixed and resonant contributions.  We focus our
attention on the spinless Falicov-Kimball model, where the problem can be
solved exactly, and the system can be tuned to go through a
Mott-Hubbard-like metal-insulator transition. Resonant effects vary in
different scattering geometries, corresponding to the symmetries of the
charge excitations scattered by the light. We do find that the Raman
response is large near the double resonance, where the transfered
frequency is close to the incident photon frequency.  We also find a joint
resonance of both the charge-transfer peak and the low-energy peak when
the incident photon frequency is on the order of the interaction strength.
In general, the resonance effects can create order of magnitude (or more)
enhancements of features in the nonresonant response, especially when the
incident photon frequency is somewhat larger than the frequency of the
nonresonant feature. Finally, we find that the resonant effects also
exhibit isosbestic behavior, even in the $A_{\textrm{1g}}$ and
$B_{\textrm{2g}}$ sectors, and it is most prominent when the incident
photon frequency is on the order of the interaction energy.
\end{abstract}


\maketitle

\section{Introduction}

Electronic Raman scattering has long been used as a direct probe of the
charge excitations of different materials. Experiments have shown a number
of interesting phenomena, especially in correlated materials.  A material
independence for Raman scattering has been seen in a number of different
correlation gap (insulating) materials ranging from FeSi\cite{Pap1}, to
SmB$_6$\cite{Pap2}, to Ca$_{3}$Ru$_{2}$O$_{7}$\cite{Pap3} to high
temperature superconductors\cite{resonance,Blumberg,irwin}
The Raman response
shows a gap opening at low temperature, but with the gap about ten times
larger than the onset temperature where the gap starts to form.  In
addition, an isosbestic point is often seen, where the Raman response at
one value of frequency is independent of temperature (at low temperature),
and curves for different temperatures appear to cross at a single point.
Resonant effects are even more interesting, as it is believed that the
resonance can cause an enhancement of the nonresonant signal by orders of
magnitude, and allow small signals to become observable.  What remains
unknown is whether these resonant enhancements dramatically change the
shape of the underlying nonresonant response.

The theoretical description of electronic Raman scattering has lagged
behind experiment. Recently, dynamical mean field theory (DMFT) has been
employed to calculate the nonresonant response in the
Falicov-Kimball\cite{paper1} and Hubbard models\cite{paper2} and to
examine inelastic x-ray scattering as well\cite{paper3}. It was found that
the theoretical calculations of the nonresonant response show much of the
behavior seen in experiment, including the large gap relative to the onset
temperature and the generic appearance of an easily observed isosbestic
point in the $B_{\textrm 1g}$ channel.

However, it is well known that many of the Raman signals in correlated
metals and insulators display complicated dependences on the incoming
photon frequency $\omega_{i}$. The resonant behavior of the $B_{1g}$
two-magnon feature at roughly $340$~meV has been well studied in the
parent insulating cuprates La$_{2}$CuO$_{4}$, YBa$_{2}$Cu$_{3}$O$_{6}$,
and Sr$_{2}$CuO$_{2}$Cl$_{2}$\cite{resonance}, where a resonance is found
for incident photon energies near 3~eV. Although recent progress has been
made\cite{Chubukov,Tohyama}, the reason for this resonance is not clear
since the resonance frequency lies above the optical absorption edge
frequency measured in the dielectric response\cite{Blumberg}, and the
photon energy is much larger than the location of the resonance peak in
the response function.

The general question of how the low-energy features (such as particle-hole
excitations near the Fermi level) and high-energy (such as charge-transfer
excitations) change under resonant conditions remains relatively
unexplored. Most treatments for Raman scattering in insulators have
focused on only the spin degrees of freedom (Heisenberg limit) in two
dimensions. For the case when the incident photon energy is much less than
the optical band gap, the Loudon-Fleury theory\cite{LF} has been widely
employed to determine resonance profiles from spin degrees of freedom via
series expansions\cite{RRP1}, exact diagonalization of small
clusters\cite{Dagotto} or quantum Monte Carlo simulations \cite{Sandvik}
of the Heisenberg model.  Modifications due to quantum
fluctuations\cite{RRP1,Sandvik}, bilayers\cite{Morr}, four-magnon
processes\cite{RRP2}, couplings beyond nearest neighbor
exchange\cite{CanaliGirvin}, and ring exchange\cite{Kampf} have all been
taken into account to give a thorough treatment of two magnon scattering
from spin degrees of freedom in the nonresonant regime. These approaches
fail when the laser frequency is tuned to lie near an optical transition.
In this regime, based on a spin-density-wave approach, Frenkel, Chubukov,
and Morr have formulated a so called ``triple-resonance'' theory from
which important features of the spectra can be derived\cite{Chubukov}.
While good agreement was obtained for the resonant profile of the two
magnon contribution to light scattering\cite{Blumberg}, general features
not related to the two-magnon peak are missed and lineshape calculations
are complex and only semiquantitative.

An approach treating the full fermionic degrees of freedom is still
lacking. Recently exact diagonalization studies of the Hubbard model have
been employed to yield lineshapes in the resonant limit from both spin and
charge degrees of freedom\cite{Tohyama}. Yet the nonresonant and mixed
terms were not taken into account. These calculations also suffer the
problems related to the finite size of the clusters (such as artificially
broadening the delta functions to approximate thermodynamic-limit
spectra). So generally, there is no theory for Raman scattering from both
charge and spin degrees of freedom which predicts spectral lineshapes
where both resonant, mixed, and nonresonant terms are treated on an equal
footing and do not suffer from finite-size effects.

In this contribution we illustrate how to calculate the full electronic
Raman response function, including contributions from the nonresonant,
mixed, and resonant processes within a single-band model.  Our model
includes interactions of the photon with all charge excitations of a
correlated fermionic system, but does not take into account any scattering
off of spin excitations. The scattering response is a complicated function
of the correlations, the temperature, the incident photon energy, and the
transfered energy. A short communication of this work has already
appeared\cite{ll8979}.

Little is known about what the mixed Raman response looks like. We find
that, as opposed to the nonresonant and resonant responses, which are
manifestly positive, the mixed response is often negative (although the
total response always remains positive). The resonant response is expected to be
large in the region where the transfered energy approaches the incident
photon energy, called the double resonance, because the energy
denominators of two pairs of the Green's functions in the bare
response function approach zero.
Interesting results are also anticipated in the strongly coupled
(Mott-insulating) regime, when the incident photon energy is close to the
interaction energy. Indeed, we find this is the case here.  We also
examine the situation where the initial photon energy is larger than the
excitation energies in the correlated band.  This is the most common
experimental situation in correlated materials with renormalized
low-energy ``bands''.  The mixed and resonant responses also behave
differently than the nonresonant response when we compare the Stokes
(energy transfered from the photon to the electrons) and the anti-Stokes
(energy transfered from the electrons to the photons) responses. These are
equal for nonresonant scattering, but the anti-Stokes response is much
smaller than the Stokes response for the mixed and resonant scattering
cases (introducing an asymmetry to the Raman scattering).

The theoretical challenge in calculating the full inelastic light
scattering response function is that the mixed diagrams involve
three-particle susceptibilities and the resonant diagrams involve
four-particle susceptibilities.  It is only in the infinite-dimensional
limit, where most of the many-particle vertex renormalizations vanish (all
three-particle and four-particle vertices do not contribute; only the
two-particle vertices enter), can one imagine performing the calculation
of these susceptibilities exactly. It turns out that because the
two-particle irreducible charge vertex is known exactly for the
Falicov-Kimball model\cite{FK}, one can calculate the full Raman response
function in this case (since the general form of the charge vertex is not
known for the Hubbard model, one can only perform approximate calculations
for that system even in infinite dimensions; nevertheless, the diagrammatic
analysis given in Section III holds for the Hubbard model, we just are
not able to evaluate the expressions).

We evaluate our exact expressions numerically and study their evolution as
functions of the incident light energy and of the transfered energy. In
the case of a correlated metal, we show how Fermi-liquid-like features
evolve as the lifetime of putative quasiparticles decreases due to
scattering.  The results are even more interesting in the correlated
insulator.  We examine what happens to the isosbestic point identified in
the nonresonant response, and how the presence of the charge gap affects
the optical scattering.

Inelastic light scattering involves a coupling of photons to electronic
charge excitations of the correlated material.  The symmetry of the
incident and scattered light relates to the symmetry of the charge
excitations that are coupled to the light. There are typically three
symmetries examined in experiments.  The $A_{\textrm{1g}}$ symmetry has
the full symmetry of the lattice. This is measured, in a system with only
nearest-neighbor hopping, by polarizing the incident and scattered light
along the diagonal direction of the hypercubic lattice, so in large
dimensions, we take the initial and final polarizations to be
$e^i=e^f=(1,1,1,...)$. The $B_{\textrm{1g}}$ symmetry is a $d$-wave-like
symmetry  that involves crossed polarizers along the diagonals.  We take
$e^i=(1,1,1,...)$ and $e^f=(-1,1,-1,1,...)$ for the $B_{\textrm{1g}}$
channel.  Finally, the $B_{\textrm{2g}}$ symmetry is another $d$-wave
symmetry rotated by 45 degrees; it requires the polarizations to satisfy
$e^i=(1,0,1,0,...)$ and $e^f=(0,1,0,1,...)$.  It turns out that the
$A_{\textrm{1g}}$ sector has contributions from nonresonant, mixed, and
resonant Raman scattering, the $B_{\textrm{1g}}$ sector has contributions
from nonresonant and resonant Raman scattering only, and the
$B_{\textrm{2g}}$ sector is purely resonant. This is generally true for a
model on a bipartite lattice with nearest-neighbor hopping only. If
longer-range hoppings are allowed, then all channels will have
nonresonant, mixed and resonant contributions.

While our approach towards analytic continuation is general, the overall
complexity of the problem limits our evaluation of the light scattering
cross section. By focusing on the spinless Falicov-Kimball model, we
present a theory of light scattering from charge degrees of freedom only;
valid for any incoming photon frequency. However, we are not able to
address scattering from spin degrees of freedom resulting in two-magnon
Raman scattering, for example. Nevertheless, we expect that our results
help frame the physics related to resonance phenomena in paramagnetic
correlated metals and insulators, and the behavior near a metal-insulator
transition.

Our plan of the paper is as follows. In Sec. II, we describe the general
analytic-continuation formula that carries one from a time-ordered
correlation function on the imaginary axis to the real response function.
The formulas are completely general, and hold for the case of inelastic
scattering of x-rays as well. The challenge is in evaluating the
corresponding response functions along the real axis, which we know how to
do only for the Falicov-Kimball model in infinite dimensions.  In Sec.
III, we evaluate the Raman scattering for the Falicov-Kimball model
explicitly, calculating all response functions, and showing in detail how
to perform all of the relevant renormalizations of the two-, three-, and
four-particle correlation functions.  In Sec. IV, we present our numerical
results for Raman scattering at half filling.  We examine the metallic
case, the insulating case, and study the evolution of the Raman response
as a function of the incident photon energy.  We present our conclusions
in Sec. V.

\section{General Analytic Continuation Formalism}

Our starting point is the expression for the inelastic light scattering
cross section derived by Shastry and Shraiman\cite{Shastry}

\begin{align}\label{SScrosssec}
    R(\bm q,\Omega) = 2\pi \sum_{i,f} \exp(-\beta\varepsilon_i)
    \delta(\varepsilon_f - \varepsilon_i - \Omega)
    \\
    \nonumber
    \times
    \left| g(\bm k_i) g(\bm k_f) e_\alpha^i e_\beta^f
    \left\langle f \left| \Hat M^{\alpha\beta}(\bm q) \right| i \right\rangle
    \right|^2 /\mathcal{Z}
\end{align}
for the scattering of electrons by photons of arbitrary wavelength (the
repeated indices $\alpha$ and $\beta$ are summed over). Here $\Omega =
\omega_i - \omega_f$ and $\bm q = \bm k_i - \bm k_f$ are the transfered
energy and momentum, respectively,  while $\omega_{i(f)}$, $\bm k_{i(f)}$,
and $\bm e^{i(f)}$ denote the energy, momentum and polarization of the
initial (final) states of the photons, $\varepsilon_{i(f)}$ refer to the
eigenstates describing the ``electronic matter'', and $g(\bm q) =
(hc^2/V\omega_{\bm q})^{1/2}$ is the ``scattering strength'' with
$\omega_{\bm q}=c|\bm q|$. Lastly, $\mathcal{Z}$ is the partition
function. For an electronic system with nearest-neighbor hopping, the
interaction with a weak external transverse electromagnetic field $\bm A$
is described by the following interacting Hamiltonian
\begin{align}\label{Hint}
    H_{\text{int}} &= -\frac{e}{\hbar c} \sum_{\bm k} \bm j(\bm k) \cdot
\bm A(-\bm k)
    \\
    \nonumber
    &+ \frac{e^2}{2\hbar^2 c^2} \sum_{\bm k \bm k'} A_\alpha(-\bm k)
    \gamma_{\alpha,\beta}(\bm k + \bm k') A_\beta(-\bm k'),
\end{align}
where
\begin{equation}\label{Hint1}
    \begin{split}
    j_\alpha(\bm q) &= \sum_{\bm k} v_\alpha(\bm k)
    c_\sigma^\dagger(\bm k + \bm q/2) c_\sigma(\bm k - \bm q/2),\\
    v_\alpha(\bm k) &= \frac{\partial\epsilon(\bm k)}{\partial k_\alpha}
    \end{split}
\end{equation}
are the current operator and Fermi velocity, respectively, and
\begin{equation}\label{Hint2}
    \gamma_{\alpha,\beta}(\bm q) = \sum_{\bm k}
    \frac{\partial^2\epsilon(\bm k)}{\partial k_\alpha\partial k_\beta}
    c_\sigma^\dagger(\bm k + \bm q/2) c_\sigma(\bm k - \bm q/2)
\end{equation}
is the so-called stress tensor. As a result, the scattering operator $\Hat
M(\bm q)$ has both nonresonant and resonant contributions
\begin{align}\label{M_oper}
    \left\langle f \left| \Hat M^{\alpha\beta}(\bm q) \right| i \right\rangle
    =&
    \left\langle f \left| \gamma_{\alpha,\beta}(\bm q) \right| i \right\rangle
    \\
    \nonumber
    +& \sum_l \left(
    \frac{\left\langle f \left| j_{\beta}(\bm k_f) \right| l \right\rangle
    \left\langle l \left| j_{\alpha}(-\bm k_i) \right| i \right\rangle}
    {\varepsilon_l - \varepsilon_i - \omega_i}
    \right.
    \\
    \nonumber
    &+
    \left.
    \frac{\left\langle f \left| j_{\alpha}(-\bm k_i) \right| l \right\rangle
    \left\langle l \left| j_{\beta}(\bm k_f) \right| i \right\rangle}
    {\varepsilon_l - \varepsilon_i + \omega_f}
    \right),
\end{align}
with the sum $l$ over intermediate states. The term with the stress tensor
is the nonresonant contribution, while the term with the square of the
current operator is the resonant contribution. Now the Raman-scattering
cross section contains nonresonant, mixed, and resonant contributions
(because it is constructed from the square of the scattering operator):
\begin{equation}\label{Raman_gen}
    R(\bm q,\Omega) = R_N(\bm q,\Omega) + R_M(\bm q,\Omega) +
                      R_R(\bm q,\Omega),
\end{equation}
where the nonresonant contribution is
\begin{align}\label{Raman_N}
    R_N&(\bm q,\Omega) = 2\pi g^2(\bm k_i) g^2(\bm k_f)
    \\
    \nonumber
    &\times
    \sum_{i,f} \frac{\exp(-\beta\varepsilon_i)}{\mathcal{Z}} \;
    \Tilde \gamma_{i,f} \;
    \Tilde \gamma_{f,i} \;
    \delta(\varepsilon_f - \varepsilon_i - \Omega),
\end{align}
the mixed contribution is
\begin{align}\label{Raman_M}
    R_M&(\bm q,\Omega) = 2\pi g^2(\bm k_i) g^2(\bm k_f) \sum_{i,f,l}
\frac{\exp(-\beta\varepsilon_i)}{\mathcal{Z}}
    \\
    \nonumber
        \times&
    \left[
    \Tilde \gamma_{i,f}
    \left(
    \frac{
    j^{(f)}_{f,l}
    j^{(i)}_{l,i}
    }
    {\varepsilon_l - \varepsilon_i - \omega_i + i0^+}
    +
    \frac{
    j^{(i)}_{f,l}
    j^{(f)}_{l,i}
    }
    {\varepsilon_l - \varepsilon_i + \omega_f - i0^+}
    \right)
    \right.
    \\
    \nonumber
    &\left.
    +
    \left(
    \frac{
    j^{(i)}_{i,l}
    j^{(f)}_{l,f}
    }
    {\varepsilon_l - \varepsilon_i - \omega_i - i0^+}
    +
    \frac{
    j^{(f)}_{i,l}
    j^{(i)}_{l,f}
    }
    {\varepsilon_l - \varepsilon_i + \omega_f + i0^+}
    \right)
    \Tilde \gamma_{f,i}
    \right]
    \\
    \nonumber
    \times &
    \delta(\varepsilon_f - \varepsilon_i - \Omega),
\end{align}
and the resonant contribution is
\begin{align}\label{Raman_R}
    R_R(\bm q,\Omega) &= 2\pi g^2(\bm k_i) g^2(\bm k_f) \sum_{i,f,l,l'}
    \frac{\exp(-\beta\varepsilon_i)}{\mathcal{Z}}
    \\
    \nonumber
    &\times\left(
    \frac{
    j^{(i)}_{i,l}
    j^{(f)}_{l,f}
    }
    {\varepsilon_l - \varepsilon_i - \omega_i - i0^+} +
    \frac{
    j^{(f)}_{i,l}
    j^{(i)}_{l,f}
    }
    {\varepsilon_l - \varepsilon_i + \omega_f + i0^+}
    \right)
    \\
    \nonumber
    &\times
    \left(
    \frac{
    j^{(f)}_{f,l'}
    j^{(i)}_{l',i}
    }
    {\varepsilon_{l'} - \varepsilon_i - \omega_i + i0^+} +
    \frac{
    j^{(i)}_{f,l'}
    j^{(f)}_{l',i}
    }
    {\varepsilon_{l'} - \varepsilon_i + \omega_f - i0^+}
    \right)
    \\
    \nonumber
    &\times
    \delta(\varepsilon_f - \varepsilon_i - \Omega).
\end{align}
In these equations, we have introduced the following symbols
\begin{equation}\label{Raman_not}
    \begin{split}
    \Tilde \gamma &= \sum_{\alpha\beta} e_\alpha^i \gamma_{\alpha,\beta}(\bm q)
    e_\beta^f, \\
    j^{(i)} &= \sum_\alpha e_\alpha^i j_\alpha(-\bm k_i), \\
    j^{(f)} &= \sum_\alpha e_\alpha^f j_\alpha(\bm k_f),
    \end{split}
\end{equation}
with the notation $A_{i,f} = \left\langle i \left| A \right| f
\right\rangle$ for the matrix elements of an operator $A$.

In general, the matrix elements that enter into
Eqs.~(\ref{Raman_N}--\ref{Raman_R}) are not easy to calculate for an
interacting system, so the summations are problematic to evaluate.
Instead, these expressions usually are evaluated via Green's function
techniques starting from correlation functions evaluated on the imaginary
axis and then performing an analytic continuation to the real axis to get
the physical response functions.  This procedure becomes more complicated
when the number of matrix elements that enter into each term in the
summations increases, because it requires the evaluation of a more
complicated correlation function on the imaginary axis. Our strategy is to
first consider the analytic continuation procedure in a general sense,
which holds for any model Hamiltonian and for arbitrary momentum transfer.
We will derive connection formulas between the Matsubara frequency axis
correlation functions and the analytically continued response functions on
the real axis.  But those expressions will require us to be able to
evaluate a number of different susceptibilities, and those expressions are
not known for arbitrary Hamiltonians.  We will show how to evaluate them
exactly for the Falicov-Kimball model in the next section.

We derive below the connection formulas between the imaginary-time
response functions and the real frequency response functions for the
general case. In all of our numerical results, we examine only optical
light scattering where we approximate $\bm k_i= \bm k_f=\bm q=0$.

\subsection{Nonresonant scattering}

The nonresonant scattering in Eq.~(\ref{Raman_N}) is proportional to the
spectral density function.  The spectral density cannot be calculated
directly but is instead obtained from the analytic continuation of the
imaginary-time response function constructed from the time-ordered product
of two stress-tensor operators
\begin{equation}
    \chi^{(2)}_{\Tilde\gamma,\Tilde\gamma}(\tau , \tau') =
    \left\langle \mathcal{T}_\tau
 \Tilde\gamma(\tau) \Tilde\gamma(\tau') \right\rangle
\end{equation}
with the $\tau$ dependence of the operator determined by the Hamiltonian
in the absence of the electromagnetic field (the symbol $\mathcal{T}_\tau$
denotes time ordering).  The first step is to calculate the double Fourier
transformation to the Matsubara frequency axis
\begin{equation}
\chi^{(2)}_{\Tilde\gamma,\Tilde\gamma}(i\nu_l, i\nu_n)=T \int_0^\beta
d\tau \int_0^\beta d\tau^\prime e^{i\nu_l\tau}
\chi^{(2)}_{\Tilde\gamma,\Tilde\gamma}(\tau,\tau^\prime)e^{i\nu_n\tau^\prime}
\end{equation}
for Bosonic Matsubara frequencies $i\nu_n=i\pi T 2n$ with $\beta=1/T$. In
thermal equilibrium, the two-particle correlation function depends only on
the difference of the two time variables, so the double Fourier transform
becomes a ``diagonal'' function, evaluated as
\begin{align}\label{chitt}
    \chi^{(2)}_{\Tilde\gamma,\Tilde\gamma}(-i\nu, i\nu) &=
    \sum_{i,f} \frac{\exp(-\beta\varepsilon_i)}{\mathcal{Z}} \frac{
    \Tilde \gamma_{i,f}
    \Tilde \gamma_{f,i}
    }
    {\varepsilon_f - \varepsilon_i - i\nu}
    \\
    \nonumber
    & \times
    \left[1-\exp(\beta(\varepsilon_i - \varepsilon_f))\right].
\end{align}
In order to extract the spectral density of states from the Matsubara
correlation function in Eq.~(\ref{chitt}), we perform the analytic
continuation $i\nu \to \Omega \pm i0^+$ which yields for the nonresonant
scattering the known expression
\begin{equation}\label{RNfin}
    R_N(\bm q,\Omega) = \frac{2\pi g^2(\bm k_i) g^2(\bm k_f)}{1-\exp(-\beta\Omega)}
        \chi_N(\bm q,\Omega),
\end{equation}
where we introduced the nonresonant response function
\begin{align}\label{chiNfin}
    \chi_N(\bm q,\Omega) = \frac1{2\pi i} & \left\{
     \chi^{(2)}_{\Tilde\gamma,\Tilde\gamma}(-\Omega-i0^+,\Omega+i0^+)
     \right.
     \\
     \nonumber
    & \left.
    - \chi^{(2)}_{\Tilde\gamma,\Tilde\gamma}(-\Omega+i0^+,\Omega-i0^+) \right\}
\end{align}
evaluated on the real axis. A similar strategy is used to determine the
mixed and resonant contributions as described in the next two subsections.

\subsection{Mixed scattering}

In the case of mixed scattering in Eq.~(\ref{Raman_M}), the calculation
begins with the multi-time correlation function constructed from the
stress tensor and two current operators
\begin{equation}\label{chitfi1}
    \chi^{(3)}_{\Tilde\gamma,f,i}(\tau , \tau' , \tau'') =
    \left\langle \mathcal{T}_\tau \Tilde\gamma(\tau) j^{(f)}(\tau') j^{(i)}(\tau'') \right\rangle.
\end{equation}
We define the Fourier transform as before, with respect to three Matsubara
frequencies (all with the same sign of the exponent).  Once again, in
thermal equilibrium we have imaginary-time-translation invariance, so the
sum of the three Matsubara frequencies must vanish, yielding
\begin{align}\label{chitfi2}
    \chi^{(3)}_{\Tilde\gamma,f,i}(i\nu_1, i\nu_2, i\nu_3) = & \delta(\nu_1+\nu_2+\nu_3)\frac{1}{\mathcal{Z}}
    \\
    \times\biggl\{\sum_{i,f,l} \Tilde\gamma_{i,f} j^{(f)}_{f,l} j^{(i)}_{l,i}
    &\left[
    \frac{\exp(-\beta\varepsilon_i)}
    {(\varepsilon_f - \varepsilon_i + i\nu_1)(\varepsilon_l - \varepsilon_i - i\nu_3)}
    \right.
    \nonumber
    \\
    &+
    \frac{\exp(-\beta\varepsilon_f)}
    {(\varepsilon_l - \varepsilon_f + i\nu_2)(\varepsilon_i - \varepsilon_f - i\nu_1)}
    \nonumber
    \\
    &+
    \left.
    \frac{\exp(-\beta\varepsilon_l)}
    {(\varepsilon_i - \varepsilon_l + i\nu_3)(\varepsilon_f - \varepsilon_l - i\nu_2)}
    \right]
    \nonumber
    \\
    + \sum_{i,f,l}  j^{(i)}_{i,l} j^{(f)}_{l,f} \Tilde\gamma_{f,i}
    &\left[
    \frac{\exp(-\beta\varepsilon_i)}
    {(\varepsilon_f - \varepsilon_i - i\nu_1)(\varepsilon_l - \varepsilon_i + i\nu_3)}
    \right.
    \nonumber
    \\
    &+
    \frac{\exp(-\beta\varepsilon_f)}
    {(\varepsilon_l - \varepsilon_f - i\nu_2)(\varepsilon_i - \varepsilon_f + i\nu_1)}
    \nonumber
    \\
    &+
    \left.
    \frac{\exp(-\beta\varepsilon_l)}
    {(\varepsilon_i - \varepsilon_l - i\nu_3)(\varepsilon_f - \varepsilon_l + i\nu_2)}
    \right]\biggr\}
    \nonumber
\end{align}
which contains $3!=6$ terms collected into two groups of terms connected
by cyclic permutations, with
\begin{equation} \label{chi3conj}
        \chi^{(3)}_{A,B,C}(i\nu_1, i\nu_2, i\nu_3) =
        \chi^{(3)}_{A^\dag, B^\dag, C^\dag }(-i\nu_1, -i\nu_2, -i\nu_3).
\end{equation}
After analytic continuation $i\nu_\alpha\to z_\alpha$ with the constraint
\begin{equation}\label{const3}
        z_1+z_2+z_3=0,
\end{equation}
one can see that the expression in Eq.~(\ref{chitfi2}) has three branch
cuts when $\Img z_\alpha\to 0^\pm$ (for $\alpha=1$, 2, or 3). Note that
the constraint in (\ref{const3}) forbids only two of the $z_\alpha$'s to
simultaneously have $\Img z_\alpha = 0^\pm$, but the imaginary part of all
three can vanish simultaneously. In order to produce the expression for
the mixed Raman cross section in Eq.~(\ref{Raman_M}), we need to focus on
the branch cuts that occur when $z_1\to -\Omega\pm i0^+$ and $z_1\to
\Omega\pm i0^+$ in order to produce the appropriate $\delta$-function and
matrix elements in the mixed scattering cross section. The corresponding
discontinuity across the branch cut when $\Img z_1 = 0$, occurs when the
terms in Eq.~(\ref{chitfi2}) are analytically continued with $z_1$ moving
onto the real axis. In the first case when $z_1 \to -\Omega \pm i0^+$, and
$z_2 \to \Omega - z_3$, we find
\begin{align}\label{chi3p1}
    &\! \frac1{2\pi i} \chi^{(3)}_{\Tilde\gamma,f,i}(z_1, z_2, z_3)
        \biggr|^{z_1\to-\Omega-i0^+}_{z_1\to-\Omega+i0^+}
    \\
    \nonumber
    &\! =
        \frac1{2\pi i}\left[ \chi^{(3)}_{\Tilde\gamma,f,i}(-\Omega-i0^+, \Omega - z_3, z_3)
        \right.
    \\
    \nonumber
    &   -
      \left.
        \chi^{(3)}_{\Tilde\gamma,f,i}(-\Omega+i0^+, \Omega - z_3, z_3) \right]
        \\
        \nonumber
    &\! =\!\left(1-e^{-\beta\Omega}\right)\!\sum_{i,f,l}\!\frac{e^{-\beta\varepsilon_i}}{\mathcal{Z}}
        \Tilde\gamma_{i,f} \left[
        \frac{j^{(f)}_{f,l}j^{(i)}_{l,i}}{\varepsilon_l - \varepsilon_i - z_3}
        +
        \frac{j^{(i)}_{f,l}j^{(f)}_{l,i}}{\varepsilon_l - \varepsilon_f + z_3}
        \right]
    \\
    \nonumber
    &\! \times\delta(\varepsilon_f - \varepsilon_i - \Omega)
\nonumber
\end{align}
and in the second case when $z_1 \to \Omega \pm i0^+$, and $z_2 \to
-\Omega - z_3$, we find
\begin{align}\label{chi3p2}
    &   \frac1{2\pi i} \chi^{(3)}_{\Tilde\gamma,f,i}(z_1, z_2, z_3)
        \biggr|^{z_1\to\Omega+i0^+}_{z_1\to\Omega-i0^+}
    \\
    \nonumber
    &\! =
        \frac1{2\pi i}\left[ \chi^{(3)}_{\Tilde\gamma,f,i}(\Omega+i0^+, -\Omega - z_3, z_3)
        \right.
    \\
    \nonumber
    &   -
      \left.
      \chi^{(3)}_{\Tilde\gamma,f,i}(\Omega-i0^+, -\Omega - z_3, z_3) \right]
        \\
        \nonumber
    &\! =\!\left(1-e^{-\beta\Omega}\right)\!\sum_{i,f,l}\frac{e^{-\beta\varepsilon_f}}{\mathcal{Z}}
        \Tilde\gamma_{i,f} \! \left[
        \frac{j^{(f)}_{f,l}j^{(i)}_{l,i}}{\varepsilon_l - \varepsilon_i - z_3}
        +
        \frac{j^{(i)}_{f,l}j^{(f)}_{l,i}}{\varepsilon_l - \varepsilon_f + z_3}
        \right]
    \\
    \nonumber
    &\! \times
        \delta(\varepsilon_f - \varepsilon_i + \Omega).
\nonumber
\end{align}
The sum on the right-hand side of Eq.~(\ref{chi3p1}), with $z_3 =
\omega_i-i0^+$, is proportional to the first two terms in
Eq.~(\ref{Raman_M}). The sum on the right-hand side of Eq.~(\ref{chi3p2}),
with an interchange of $i \leftrightarrow f$ in the summation and $z_3 =
-\omega_i-i0^+$, is proportional to the last two terms in (\ref{Raman_M}).
Hence, we arrive at the general expression for the mixed scattering
\begin{equation}\label{RMfin}
    R_M(\bm q,\Omega) = \frac{2\pi g^2(\bm k_i) g^2(\bm k_f)}{1-\exp(-\beta\Omega)}
        \chi_M(\bm q,\Omega)
\end{equation}
with the mixed Raman response function defined by
\begin{align} \label{chiMfin}
    \chi_M(\bm q,\Omega) =
    & \frac1{2\pi i} \Bigl[
         \chi^{(3)}_{\Tilde\gamma,f,i}(-\Omega-i0^{+},-\omega_f + i0^{+},\omega_i - i0^{+})
        \\
        \nonumber
        & - \chi^{(3)}_{\Tilde\gamma,f,i}(-\Omega+i0^{+},-\omega_f + i0^{+},\omega_i - i0^{+})
         \\
         \nonumber
        & + \chi^{(3)}_{\Tilde\gamma,f,i}(\Omega+i0^{+},\omega_f + i0^{+},-\omega_i - i0^{+})
        \\
        \nonumber
        & - \chi^{(3)}_{\Tilde\gamma,f,i}(\Omega-i0^{+},\omega_f + i0^{+},-\omega_i - i0^{+})
    \Bigr]
\end{align}
on the real axis. The operators $\Tilde\gamma(\bm q)$, $j^{(f)}(\bm k_f)$
and $j^{(i)}(\bm k_i)$ are Hermitian for optical light scattering, which
has vanishing momentum $\bm k_i=\bm k_f=\bm q=0$, and, with the use of
Eq.~(\ref{chi3conj}), the expression in Eq.~(\ref{chiMfin}) can be
rewritten as:
\begin{align} \label{chiMfin2}
    \chi_M(\Omega) = & \chi_M(\bm q=0,\Omega) 
         \\
         \nonumber
    =&  \frac1{2\pi i}  \Bigl[
         \chi^{(3)}_{\Tilde\gamma,f,i}(\Omega+i0^{+},\omega_f - i0^{+},-\omega_i + i0^{+})
         \\
         \nonumber
         & - \chi^{(3)}_{\Tilde\gamma,f,i}(\Omega-i0^{+},\omega_f - i0^{+},-\omega_i + i0^{+})
         \\
         \nonumber
        & + \chi^{(3)}_{\Tilde\gamma,f,i}(\Omega+i0^{+},\omega_f + i0^{+},-\omega_i - i0^{+})
         \\
         \nonumber
        & - \chi^{(3)}_{\Tilde\gamma,f,i}(\Omega-i0^{+},\omega_f + i0^{+},-\omega_i - i0^{+})
    \Bigr].
\end{align}

\subsection{Resonant scattering}

For the resonant scattering case in Eq.~(\ref{Raman_R}), the procedure is
similar: one has to calculate the multi-time correlation function
constructed from the four current operators
\begin{equation}\label{chiiffi1}
    \chi^{(4)}_{i,f,f,i}(\tau_1, \tau_2 , \tau_3 , \tau_4) {=}
    \left\langle\! \mathcal{T}_\tau j^{(i)}(\tau_1) j^{(f)}(\tau_2)
    j^{(f)}(\tau_3) j^{(i)}(\tau_4) \!\right\rangle.
\end{equation}
Once again, defining the Fourier transform in terms of four Matsubara
frequencies (with the same sign in the exponent) yields the following
result (with the delta function arising from the time-translation
invariance)
\begin{align}\label{chiiffi2}
 &   \chi^{(4)}_{i,f,f,i}(i\nu_1, i\nu_2, i\nu_3, i\nu_4) = \delta(\nu_1+\nu_2+\nu_3+\nu_4)
    \\
    \nonumber
    &\times
    \left[
    \Tilde\chi^{(4)}_{i,f,f,i}(i\nu_1, i\nu_2, i\nu_3, i\nu_4)
    \right.
    +
    \Tilde\chi^{(4)}_{i,f,i,f}(i\nu_1, i\nu_2, i\nu_4, i\nu_3)
    \\
    \nonumber
    &
        +
        \left.
    \Tilde\chi^{(4)}_{i,i,f,f}(i\nu_1, i\nu_4, i\nu_2, i\nu_3)
    \right].
\end{align}
Here we introduce the generic four-particle susceptibility
%\begin{widetext}
\begin{align}\label{chiiffi3}
    \Tilde\chi^{(4)}_{A,B,C,D}(i\nu_1, i\nu_2, i\nu_3, i\nu_4) =
    \sum_{i,f,l,l'} A_{i,l} B_{l,f} C_{f,l'} D_{l',i} \frac{1}{\mathcal{Z}}
    \\
    \nonumber
    \times
    \left[
    \frac{\exp(-\beta\varepsilon_i)}
    {(\varepsilon_l - \varepsilon_i + i\nu_1)(\varepsilon_{l'} - \varepsilon_i - i\nu_4)
     (\varepsilon_f - \varepsilon_i - i\nu_3 - i\nu_4)} \right.
     \\
    \nonumber
     +
    \frac{\exp(-\beta\varepsilon_l)}
    {(\varepsilon_f - \varepsilon_l + i\nu_2)(\varepsilon_i - \varepsilon_l - i\nu_1)
     (\varepsilon_{l'} - \varepsilon_l - i\nu_4 - i\nu_1)}
     \\
    \nonumber
     +
    \frac{\exp(-\beta\varepsilon_f)}
    {(\varepsilon_{l'} - \varepsilon_f + i\nu_3)(\varepsilon_l - \varepsilon_f - i\nu_2)
     (\varepsilon_i - \varepsilon_f - i\nu_1 - i\nu_2)}
     \\
    \nonumber
     +\left.
    \frac{\exp(-\beta\varepsilon_{l'})}
    {(\varepsilon_i - \varepsilon_{l'} + i\nu_4)(\varepsilon_f - \varepsilon_{l'} - i\nu_3)
     (\varepsilon_f - \varepsilon_{l'} - i\nu_2 - i\nu_3)}
    \right]
    \\
    \nonumber
    +\sum_{i,f,l,l'} D_{i,l'} C_{l',f} B_{f,l} A_{l,i} \frac{1}{\mathcal{Z}}
    \\
    \nonumber
    \times
    \left[
    \frac{\exp(-\beta\varepsilon_i)}
    {(\varepsilon_l - \varepsilon_i - i\nu_1)(\varepsilon_{l'} - \varepsilon_i + i\nu_4)
     (\varepsilon_f - \varepsilon_i + i\nu_3 + i\nu_4)} \right.
     \\
    \nonumber
     +
    \frac{\exp(-\beta\varepsilon_l)}
    {(\varepsilon_f - \varepsilon_l - i\nu_2)(\varepsilon_i - \varepsilon_l + i\nu_1)
     (\varepsilon_{l'} - \varepsilon_l + i\nu_4 + i\nu_1)}
     \\
    \nonumber
     +
    \frac{\exp(-\beta\varepsilon_f)}
    {(\varepsilon_{l'} - \varepsilon_f - i\nu_3)(\varepsilon_l - \varepsilon_f + i\nu_2)
     (\varepsilon_i - \varepsilon_f + i\nu_1 + i\nu_2)}
     \\
    \nonumber
     +\left.
    \frac{\exp(-\beta\varepsilon_{l'})}
    {(\varepsilon_i - \varepsilon_{l'} - i\nu_4)(\varepsilon_f - \varepsilon_{l'} + i\nu_3)
     (\varepsilon_f - \varepsilon_{l'} + i\nu_2 + i\nu_3)}
    \right]
\end{align}
%\end{widetext}
with
\begin{align}\label{chiiffi4}
    \Tilde\chi^{(4)}_{A,B,C,D}&(i\nu_1, i\nu_2, i\nu_3, i\nu_4)
    \\
    \nonumber
    &=\Tilde\chi^{(4)}_{A^\dag,B^\dag,C^\dag,D^\dag }(-i\nu_1, -i\nu_2, -i\nu_3, -i\nu_4).
\end{align}
The expression in Eq.~(\ref{chiiffi2}) contains $4!=24$ terms collected
into six different groups of the terms, with each group member connected
by the cyclic permutation of four objects.

After analytic continuation $i\nu_\alpha\to z_\alpha$ with the constraint
\begin{equation}
        z_1+z_2+z_3+z_4=0,
\end{equation}
one can see that the expression in Eq.~(\ref{chiiffi2}) has branch cuts
when any $\Img z_\alpha\to 0^\pm$ or when any pair $\Img
(z_\alpha+z_\beta)\to 0^\pm$. The $\delta$-function in the expression for
the resonant scattering cross section in Eq.~(\ref{Raman_R}) is connected
to the branch cut at $z_3+z_4=-z_1-z_2\to \Omega\pm i0^+$ and the
discontinuity of the response function across this branch cut is equal to
\begin{align}\label{chi4p1}
        \frac1{2\pi i} \chi^{(4)}_{i,f,f,i}(z_1, z_2, z_3, z_4)
        \biggr|^{z_3+z_4=-z_1-z_2\to \Omega+i0^+}_{z_3+z_4=-z_1-z_2\to \Omega-i0^+}
    \\
   \nonumber
        =\left(1-e^{-\beta\Omega}\right)\sum_{i,f,l,l'}\frac{e^{-\beta\varepsilon_i}}{\mathcal{Z}}
        \delta(\varepsilon_f - \varepsilon_i - \Omega)
        \\
        \nonumber
        \times
        \left[
        \frac{j^{(i)}_{i,l}j^{(f)}_{l,f}j^{(f)}_{f,l'}j^{(i)}_{l'i}}
        {(\varepsilon_l - \varepsilon_i + z_1)(\varepsilon_{l'} - \varepsilon_i - \Omega + z_3)}
        \right.
    \\
    \nonumber
        +
        \frac{j^{(f)}_{i,l}j^{(i)}_{l,f}j^{(i)}_{f,l'}j^{(f)}_{l',i}}
        {(\varepsilon_l - \varepsilon_i - \Omega - z_1)(\varepsilon_{l'} - \varepsilon_i - z_3)}
        \\
        \nonumber
        +
        \frac{j^{(i)}_{i,l}j^{(f)}_{l,f}j^{(i)}_{f,l'}j^{(f)}_{l',i}}
        {(\varepsilon_l - \varepsilon_i + z_1)(\varepsilon_{l'} - \varepsilon_i - z_3)}
        \\
        \nonumber
        +
        \left.
        \frac{j^{(f)}_{i,l}j^{(i)}_{l,f}j^{(f)}_{f,l'}j^{(i)}_{l',i}}
        {(\varepsilon_l - \varepsilon_i - \Omega - z_1)
         (\varepsilon_{l'} - \varepsilon_i - \Omega + z_3)}
        \right] .
\end{align}
The analytic continuation procedure then requires us to take the following
limits
\begin{align}\label{chi4p2}
        z_1 &\to - \omega_i - i0^+,
        \\ \nonumber
        z_2 &\to \omega_f + i0^+,
        \\ \nonumber
        z_3 &\to - \omega'_f + i0^+,
        \\ \nonumber
        z_4 &\to \omega'_i - i0^+
\end{align}
and then take the limit
\begin{equation}
        \omega'_i-\omega_i=\omega'_f-\omega_f \to 0
\end{equation}
in order to reproduce an expression proportional to the resonant
scattering cross section in Eq.~(\ref{Raman_R}). The final general
expression for the resonant scattering becomes
\begin{equation}\label{RRfin}
    R_R(\bm q,\Omega) = \frac{2\pi g^2(\bm k_i) g^2(\bm k_f)}{1-\exp(-\beta\Omega)}
        \chi_R(\bm q,\Omega)
\end{equation}
with the resonant Raman response function defined by
\begin{widetext}
\begin{align} \label{chiRfin}
    \chi_R(\bm q,\Omega) =
    \frac1{2\pi i}
    \left.\left\{
    \chi^{(4)}_{i,f,f,i}(z_1, z_2, z_3, z_4)
        \biggr|^{z_3+z_4=-z_1-z_2\to \Omega+i0^+}_{z_3+z_4=-z_1-z_2\to \Omega-i0^+}
        \right\}
        \right|_{\left.\substack{
        z_1 \to - \omega_i - i0^+\\
        z_2 \to \omega_f + i0^+\\
        z_3 \to - \omega'_f + i0^+\\
        z_4 \to \omega'_i - i0^+
        }
        \right|_{\substack{
        \omega'_i - \omega_i \to 0\\
        \omega'_f - \omega_f \to 0
        }}};
\end{align}
\end{widetext}
note that it is critical to perform the analytic continuation of
$z_3+z_4=-z_1-z_2\to \Omega\pm i0^+$ first and then analytically continue
the other frequencies [as in Eq.~(\ref{chi4p1}) and (\ref{chi4p2})] since
these procedures do not commute with one another.

\section{Exact Results for the Falicov-Kimball model}

We now evaluate the general expressions derived above for the case of
optical Raman scattering, where all momenta vanish ($\bm k_i=\bm k_f =\bm
q=0$) and for the spinless Falicov-Kimball model.  The Falicov-Kimball
model involves the interaction of conduction electrons with localized
electrons and has the following Hamiltonian\cite{FK}
\begin{align}
 H=&-\frac{t^*}{2\sqrt{D}}\sum_{\langle
i,j\rangle}(c^\dagger_ic_j+c^\dagger_jc_i) + E_{f}\sum_{i}w_{i}
\nonumber\\
&-\mu\sum_{i}(c^{\dagger}_{i}c_{i}+w_{i}) +U\sum_ic^\dagger_ic_iw_i
\label{eq: hamiltonian}
\end{align}
where $c^\dagger_i$ ($c_i$) create (destroy) a conduction electron at site
$i$, $w_i$ is a classical variable (representing the localized electron
number at site $i$) that equals 0 or 1, $t^*$ is a renormalized hopping
matrix that is nonzero between nearest neighbors on a hypercubic lattice
in $D$-dimensions (and we take the limit $D\rightarrow\infty$), and $U$ is
the local screened Coulomb interaction between conduction and localized
electrons. $\langle i,j\rangle$ denotes a sum over sites $i$ and nearest
neighbors $j$. $E_{f}$ and $\mu$ are adjusted to set the average filling
of conduction and localized electrons. In our calculations the average
filling for each is set to $1/2$, respectively ($\mu=U/2,~E_f=0$).

This model can be solved exactly by using DMFT, as first described by
Brandt and Mielsch\cite{BM}.  The algorithm used to solve for the local
Green's function at site $i$, defined by
\begin{equation}
G_i(\tau)=-\langle \mathcal{T}_\tau c_i(\tau)c^\dagger_i(0)\rangle
\label{eq: greendef}
\end{equation}
where the angle brackets denote the trace weighted by the Boltzmann factor
$\exp[-\beta H]/\mathcal{Z}$. We usually work with the Fourier transform
of the imaginary-time Green's function to yield the Matsubara frequency
Green's function. The momentum-dependent Green's function becomes
\begin{equation}\label{G_DMFT}
        G_{m}({\bm k}) = \frac1{Z_m-\epsilon_{\bm k}},
\end{equation}
with
\begin{equation}
        \epsilon_{\bm k} = -\lim_{D\rightarrow\infty}
\frac{t^*}{\sqrt{D}}\sum_{\alpha=1}^D \cos k_\alpha
\end{equation}
being the noninteracting band energy, and
\begin{equation}
        Z_m = i\omega_m + \mu - \Sigma_m.
\end{equation}
The local self-energy $\Sigma_m$ is a solution of the following set of
equations:
\begin{align}
        G_m=\frac1N \sum_{\bm k} \frac1{Z_m-\epsilon_{\bm k}} = \frac1{Z_m-\lambda_m}
        \\
        \nonumber
        = \frac{w_1}{i\omega_m+\mu-\lambda_m-U} +
        \frac{1-w_1}{i\omega_m+\mu-\lambda_m},
\end{align}
where we introduced the self-consistent dynamical mean-field of Brandt and
Mielsch (denoted $\lambda$); the self-energy  can be expressed as a simple
function of this field
\begin{equation}
        \Sigma_m = U w_1+\frac{U^2 w_1 (1-w_1)}{i\omega_m+\mu-\lambda_m-U(1-w_1)}.
\end{equation}
Here $w_{1}$ is given by
$w_{1}=e^{[-\beta(E_{f}-\mu)]}\mathcal{Z}_{0}(U-\mu)/\mathcal{Z}$, with
$\mathcal{Z}_{0}(\mu)=2e^{\beta\mu/2}\prod_{n=-\infty}^{\infty}(i\omega_{n}-\mu-\lambda_{n})/i\omega_{n}$.

\subsection{Nonresonant scattering}

\begin{figure}[!htbf]
        \begin{center}
                \includegraphics[width=0.70\columnwidth]{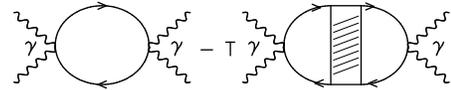}
        \end{center}
        \caption{Feynman diagrams for nonresonant Raman scattering. The
wavy lines denote photon propagators and the solid lines denote electron
propagators.  The cross-hatched rectangle is the \textit{reducible} charge
vertex.  In the $B_{\textrm{1g}}$ channel, only the bare (first) diagram
enters, while in the $A_{\textrm{1g}}$ channel both diagrams enter. The
symbol $\gamma$ denotes the stress-tensor vertex of the corresponding
electron-photon interaction.}
        \label{fig:nonresonant_diagram}
\end{figure}
The case of nonresonant Raman scattering was considered by two of
us\cite{paper1}, so we only sketch the derivation to show our notation and
to present the final results. In general, the two-time correlation
function in Eq.~(\ref{chitt}), constructed from the stress operators, can
be represented by the summation of a generalized ``polarization''
\begin{equation}\label{chi2nr}
        \chi^{(2)}_{\Tilde\gamma,\Tilde\gamma}(-i\nu, i\nu) =
        T\sum_m \Pi\left(i\omega_m,i\omega_{m+\nu}\right),
\end{equation}
where we use a shorthand notation $i\omega_{m+\nu}=i\omega_m+i\nu$ and
$G_{m+\nu} = G(i\omega_{m+\nu}) = G(i\omega_m+i\nu)$ and similarly for
$\Sigma$ and $Z$. A tedious calculation shows that\cite{paper1}
\begin{equation}\label{Pi2B1g}
        \Pi\left(i\omega_m,i\omega_{m+\nu}\right) = - \frac{t^{*2}}2
        \frac{G_m-G_{m+\nu}}{Z_{m+\nu}-Z_m}
\end{equation}
in the $B_{1g}$ channel and
\begin{equation}\label{Pi2A1g}
        \Pi\left(i\omega_m,i\omega_{m+\nu}\right) = - \frac1{i\nu}
        \frac{\Sigma_m-\Sigma_{m+\nu}}{G^{-1}_{m+\nu}-G^{-1}_m}(Z_{m+\nu}-Z_m)
\end{equation}
in the $A_{1g}$ channel (see Fig.~\ref{fig:nonresonant_diagram} for the
relevant Feynman diagrams).

Since one can show that there are no additional singularities or
non-analyticities in Eqs.~(\ref{Pi2B1g}) and (\ref{Pi2A1g}) connected with
the denominators,\cite{SFM} one can directly perform the analytic
continuation and replace the sum over Matsubara frequencies in
Eq.~(\ref{chi2nr}) by an integral over the real axis
\begin{align}\label{chi2nr2}
        \chi^{(2)}_{\Tilde\gamma,\Tilde\gamma}&(-i\nu, i\nu) =
        \frac1{2\pi i} \int^{+\infty}_{-\infty} d\omega f(\omega)
        \\
        \nonumber
        &\times
        \left[ \Pi(\omega-i0^+,\omega+i\nu)
        - \Pi(\omega+i0^+,\omega+i\nu)
        \right.
        \\
        \nonumber
        &+
        \left.
        \Pi(\omega-i\nu,\omega-i0^+)
        - \Pi(\omega-i\nu,\omega+i0^+) \right],
\end{align}
where $f(\omega)=1/[1+\exp(\beta\omega)]$ is the Fermi distribution
function. After substituting Eq.~(\ref{chi2nr2}) into the expression for
the non-resonant response function in Eq.~(\ref{chiNfin}), we obtain
\begin{align}\label{RNfinFK}
    \chi_N(\Omega) =
    \frac2{(2\pi i)^2} \int^{+\infty}_{-\infty} d\omega [f(\omega)-f(\omega+\Omega)]
        \\
        \nonumber
        \times
    \Real\left\{ \Pi(\omega-i0^+,\omega+\Omega+i0^+)
    \right.
        \\
        \nonumber
        \left.
    - \Pi(\omega-i0^+,\omega+\Omega-i0^+)
      \right\} .
\end{align}
Now we can take the trivial analytic continuation of Eqs.~(\ref{Pi2B1g})
and (\ref{Pi2A1g}) to find the final expressions for the nonresonant Raman
response function:
\begin{align}\label{RNfinFKB1g}
    \chi_{N,B_{\textrm{1g}}}&(\Omega) =
    \frac{t^{*2}}{4\pi^2} \int^{+\infty}_{-\infty} d\omega [f(\omega)-f(\omega+\Omega)]
        \\
        \nonumber
        &\times
    \Real\left\{
    \frac{G(\omega)-G^*(\omega+\Omega)}{Z^*(\omega+\Omega)-Z(\omega)}
    -
    \frac{G(\omega)-G(\omega+\Omega)}{Z(\omega+\Omega)-Z(\omega)}
      \right\}
\end{align}
in the $B_{1g}$ channel, and
\begin{align}\label{RNfinFKA1g}
    \chi_{N,A_{\textrm{1g}}}(\Omega)& =
    \frac1{2\pi^2\Omega} \int^{+\infty}_{-\infty} d\omega [f(\omega)-f(\omega+\Omega)]
        \\
        \nonumber
        \times
    \Real
    &\left\{
    \frac{[\Sigma(\omega)-\Sigma^*(\omega+\Omega)][Z^*(\omega+\Omega)-Z(\omega)]}
    {G^{-1*}(\omega+\Omega)-G^{-1}(\omega)}
    \right.
    \\
    \nonumber
    &-
    \left.
    \frac{[\Sigma(\omega)-\Sigma(\omega+\Omega)][Z(\omega+\Omega)-Z(\omega)]}
    {G^{-1}(\omega+\Omega)-G^{-1}(\omega)}
    \right\}
\end{align}
in the $A_{1g}$ channel, respectively.

\subsection{Mixed scattering}

\begin{figure}[!htbf]
    \begin{center}
        \includegraphics[width=0.70\columnwidth]{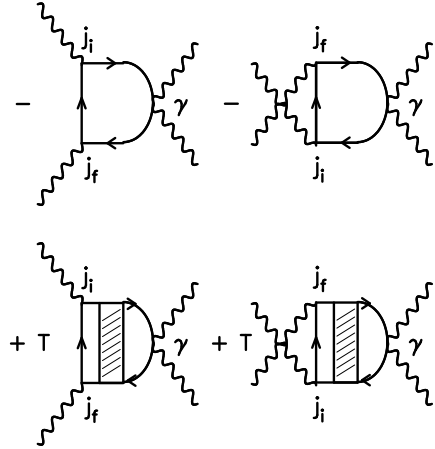}
    \end{center}
    \caption{Feynman diagrams for the mixed contributions to
Raman scattering. The symbols $j_f$ and $j_i$ remind us to include the
relevant vertex factors from the current operator in the electron-photon
interaction. The mixed contribution vanishes in the $B_{\textrm{2g}}$
channel, it consists of only the bare diagrams on the top line in the
$B_{\textrm{1g}}$ channel (and turns out to be a $1/D$ correction), and
all diagrams enter for the $A_{\textrm{1g}}$ channel.}
    \label{fig:mixed_diagram}
\end{figure}

The mixed Raman response corresponds to the scattering processes that
involve three external vertices: one stress tensor and two current
operators, and there are two types of diagrams corresponding to the direct
and exchange processes (see Fig.~\ref{fig:mixed_diagram}). There is no
mixed Raman response for the $B_{2g}$ channel because the stress tensor
vanishes for the case of nearest neighbor hopping only. In the $B_{1g}$
channel it appears to be only a bare response (we will see below that it
actually vanishes) and for the $A_{1g}$ channel, the bare mixed response
is renormalized by the irreducible charge vertex.

\subsubsection{$B_{1g}$ channel}

In the $B_{1g}$ channel, the mixed Raman response contains only the bare
direct and exchange contributions (first two terms in
Fig.~\ref{fig:mixed_diagram}, respectively):
\begin{align}\label{chi3B1g}
    \chi^{(3)}_{\Tilde\gamma,f,i}(i\nu_i-i\nu_f, i\nu_f, -i\nu_i) =
    T\sum_m \frac1N\sum_{\bm k} \frac{t^{*3}}{D^{3/2}}
        \\
        \nonumber
        \times
    \sum_{\alpha=1}^D (-1)^\alpha \cos k_\alpha
    \sum_{\beta=1}^D (-1)^\beta \sin k_\beta
    \sum_{\gamma=1}^D \sin k_\gamma
    \\
    \nonumber
    \times \left[G_{m}({\bm k})G_{m-\nu_f}({\bm k})G_{m+\nu_i-\nu_f}({\bm k})
    \right.
        \\
        \nonumber
        \left.
    +G_{m}({\bm k})G_{m+\nu_f}({\bm k})G_{m-\nu_i+\nu_f}({\bm k})\right],
\end{align}
because the symmetry of all two and three-particle vertices is that of the
lattice ($A_{\textrm{1g}}$), so all renormalizations vanish (recall the
current operator has odd parity, whereas $\epsilon_{\bm k}$ is even in
$\bm k$). The expression in Eq.~(\ref{chi3B1g}) has nonzero values only
when the subscripts are equal $\alpha=\beta=\gamma$. In this case, we
expand the product of Green's functions into partial fractions over
$\epsilon_{\bm k}$ and the summations over momentum involve only
expressions of the type [$\zeta_m=-\sgn (\Img Z_m)$]:
\begin{align}\label{sum_k1}
    \frac1N\sum_{\bm k} \frac{t^{*3}}{D^{3/2}}
    \sum_{\alpha=1}^D \cos k_\alpha \sin^2 k_\alpha
    \frac1{Z_m-\epsilon_{\bm k}}
        \\
        \nonumber
    =
    \frac{it^{*3}}{\sqrt D} \int_0^{\zeta_m\infty} d\lambda\; e^{-i\lambda Z_m}
    J_0^{D-1}\left(\frac{\lambda t^*}{\sqrt D}\right)
        \\
        \nonumber
        \times
    \int_{-\pi}^{\pi} \frac{dk}{2\pi} \sin^2 k \cos k\; e^{i\frac{\lambda t^*}{\sqrt D}\cos k}
\end{align}
with $J_0$ being Bessel's function. The last  exponent is expanded in a
power series over $\frac{\lambda t^*}{\sqrt D}$ that yields in the
$D\to\infty$ limit
\begin{align}
        \lim_{D\rightarrow\infty}-\frac{t^{*4}}{8D}\frac{d}{dZ_m}
        \left[
        i\zeta_m\frac{\sqrt\pi}{t^*} e^{-{Z_m^2}/{t^{*2}}} \erfc \left(i\zeta_m\frac{Z_m}{t^*}\right)
        \right]
        \\
        \nonumber
        =
        \lim_{D\rightarrow\infty}\frac{t^{*2}}{4D}\left(Z_m G_m - 1\right)\to 0
\end{align}
so the mixed contribution vanishes in the $B_{1g}$ channel.

\subsubsection{$A_{1g}$ channel}

In the $A_{1g}$ channel, the mixed Raman response contains both bare and
renormalized contributions:
\begin{align}\label{chi3A1g}
    \chi^{(3)}_{\Tilde\gamma,f,i}&(i\nu_i-i\nu_f, i\nu_f, -i\nu_i) =\\
\nonumber
    T&\sum_m \Biggr\{
    -\frac1N\sum_{\bm k} \epsilon_{\bm k}
    \frac{t^{*2}}{D}
    \sum_{\alpha=1}^D \sin k_\alpha
    \sum_{\beta=1}^D \sin k_\beta
\\
\nonumber \times&
    G_{m}({\bm k})G_{m-\nu_f}({\bm k})G_{m+\nu_i-\nu_f}({\bm k})
    \\
    \nonumber
    -&
    \frac1N\sum_{\bm k} \frac{t^{*2}}{D}
    \sum_{\alpha,\beta=1}^D \sin k_\alpha
    \sin k_\beta \\
\nonumber \times&
    G_{m}({\bm k})G_{m-\nu_f}({\bm k})G_{m+\nu_i-\nu_f}({\bm k})
        \\
        \nonumber
        \times&
        T
    \Tilde\Gamma\left(i\omega_m, i\omega_m+i\nu_i-i\nu_f \right)
        \\
        \nonumber
        \times&
    \frac1N\sum_{\bm k} \epsilon_{\bm k}
    G_{m}({\bm k})G_{m+\nu_i-\nu_f}({\bm k})\Biggr\}
    \\
    \nonumber
  & +\begin{bmatrix}i\nu_i\to-i\nu_i \\ i\nu_f\to-i\nu_f \end{bmatrix}.
\end{align}
The renormalizations are only with respect to two-particle vertices,
because the current operators are odd in parity and cannot be renormalized
by a local three-particle vertex (note that we cannot provide a general
proof that the
relevant three-particle vertex is local, but a strong-coupling analysis
indicates this is so to lowest order). Here
\begin{align}\label{BSvrtx}
    &   \Tilde\Gamma\left(i\omega_m, i\omega_m+i\nu_i-i\nu_f\right)
        \\
        \nonumber
    &   = \frac {\Gamma\left(i\omega_m, i\omega_m+i\nu_i-i\nu_f\right)}
        {1-T\Gamma\left(i\omega_m, i\omega_{m+\nu_i-\nu_f}\right)
        \frac1N\sum_{\bm k} G_{m}({\bm k})G_{m+\nu_i-\nu_f}({\bm k})}
\end{align}
is the total (reducible) charge vertex. In the $D=\infty$ Falicov-Kimball
model, the irreducible charge vertex satisfies
\begin{equation}
        \Gamma\left(i\omega_m, i\omega_m+i\nu\right)
        = \frac1T \frac{\Sigma_m - \Sigma_{m+\nu}}{G_m - G_{m+\nu}}
\end{equation}
on the Matsubara frequency axis\cite{SFM}. Substituting into the
expression for the reducible charge vertex gives
\begin{align}\label{gamma}
        \Tilde\Gamma\left(i\omega_m, i\omega_{m+\nu_i-\nu_f}\right)
        = \frac1T \frac {Z_{m+\nu_i-\nu_f}{-}Z_m}{i\nu_i-i\nu_f}
        \frac{\Sigma_m{-}\Sigma_{m+\nu_i-\nu_f}}{G_m{-}G_{m+\nu_i-\nu_f}}
        \\
        = \frac1T
        \left(
        \frac{Z_m{-}Z_{m+\nu_i-\nu_f}}{G_m{-}G_{m+\nu_i-\nu_f}}
        +
        \frac{\left(Z_{m+\nu_i-\nu_f}-Z_m\right)^2}{(i\nu_i-i\nu_f)(G_m{-}G_{m+\nu_i-\nu_f})}
        \right) .
        \nonumber
\end{align}

Now Eq.~(\ref{chi3A1g}) has nonzero values only when $\alpha=\beta$ and,
noting that in the $D\to \infty$ limit one can replace $\sin^2 k_\alpha$
by its average value $\frac12$, yields
\begin{align}\label{chi3m}
    \chi^{(3)}_{\Tilde\gamma,f,i}&(i\nu_i-i\nu_f, i\nu_f, -i\nu_i)
    \\
    \nonumber
    = &
    T \sum_m \left[
    \Pi^{(3)}(i\omega_m-i\nu_f,i\omega_m+i\nu_i-i\nu_f,i\omega_m)
    \right.
    \\
    \nonumber
    &\left.
    +\Pi^{(3)}(i\omega_m+i\nu_i,i\omega_m+i\nu_i-i\nu_f,i\omega_m)
    \right],
\end{align}
where
\begin{align}\label{Pi3A1g}
 &   \Pi^{(3)}(i\omega_m-i\nu_f,i\omega_m+i\nu_i-i\nu_f,i\omega_m)
    \\
    \nonumber
  &  =
    \frac{t^{*2}}{2(i\nu_i-i\nu_f)}
        \frac{\Sigma_m-\Sigma_{m+\nu_i-\nu_f}}{G_m-G_{m+\nu_i-\nu_f}}
        \\
        \nonumber
    &   \times\!
        \left[
        G_{m+\nu_i-\nu_f}\frac{G_m-G_{m-\nu_f}}{Z_{m-\nu_f}-Z_m} -
        G_{m}\frac{G_{m+\nu_i-\nu_f}-G_{m-\nu_f}}{Z_{m-\nu_f}-Z_{m+\nu_i-\nu_f}}
        \right].
\end{align}
In the case when there are neither singularities nor non-analyticities in
Eq.~(\ref{Pi3A1g}) connected with the denominators, one can trivially
perform the analytic continuation and replace the sum over Matsubara
frequencies in Eq.~(\ref{chi3m}) by an integral over the real axis:
\begin{align}\label{chi3mr}
    &\chi^{(3)}_{\Tilde\gamma,f,i}(i\nu_i-i\nu_f, i\nu_f, -i\nu_i) =
        \frac1{2\pi i} \int^{+\infty}_{-\infty} d\omega f(\omega)
        \\
        \nonumber
    \times&\left[
    \Pi^{(3)}(\omega-i\nu_f,\omega+i\nu_i-i\nu_f,\omega-i0^+)
    \right.
        \\
        \nonumber
   & -
    \Pi^{(3)}(\omega-i\nu_f,\omega+i\nu_i-i\nu_f,\omega+i0^+)
    \\
    \nonumber
    &+
    \Pi^{(3)}(\omega-i\nu_i,\omega-i0^+,\omega-i\nu_i+i\nu_f)
        \\
        \nonumber
   & -
    \Pi^{(3)}(\omega-i\nu_i,\omega+i0^+,\omega-i\nu_i+i\nu_f)
    \\
    \nonumber
    &+
    \Pi^{(3)}(\omega+i\nu_f,\omega-i\nu_i+i\nu_f,\omega-i0^+)
        \\
        \nonumber
   & -
    \Pi^{(3)}(\omega+i\nu_f,\omega-i\nu_i+i\nu_f,\omega+i0^+)
    \\
    \nonumber
    &+
    \Pi^{(3)}(\omega+i\nu_i,\omega-i0^+,\omega+i\nu_i-i\nu_f)
        \\
        \nonumber
   & -
    \Pi^{(3)}(\omega+i\nu_i,\omega+i0^+,\omega+i\nu_i-i\nu_f)
    \\
    \nonumber
    &+
    \Pi^{(3)}(\omega-i0^+,\omega+i\nu_i,\omega+i\nu_f)
        \\
        \nonumber
   & -
    \Pi^{(3)}(\omega+i0^+,\omega+i\nu_i,\omega+i\nu_f)
    \\
    \nonumber
    &+
    \Pi^{(3)}(\omega-i0^+,\omega-i\nu_i,\omega-i\nu_f)
        \\
        \nonumber
   & -\left.
    \Pi^{(3)}(\omega+i0^+,\omega-i\nu_i,\omega-i\nu_f)
    \right].
\end{align}
Only the first eight terms contain the difference $i\nu_i-i\nu_f$ and
hence contribute to the mixed scattering. Substituting Eq.~(\ref{chi3mr})
into Eq.~(\ref{chiMfin2}) we get the final expression:
\begin{align}\label{RMfinFK}
    \chi_{M,A_{\textrm{1g}}}&(\Omega) =
    \frac2{(2\pi i)^2} \int^{+\infty}_{-\infty} d\omega
    [f(\omega)-f(\omega+\Omega)]
    \\
    \nonumber
    \times\Real & \left\{
    \Pi^{(3)}(\omega-\omega_f+i0^+,\omega+\Omega+i0^+,\omega-i0^+)
    \right.
    \\
    \nonumber
   & -
    \Pi^{(3)}(\omega-\omega_f+i0^+,\omega+\Omega-i0^+,\omega-i0^+)
    \\
    \nonumber
    &+
    \Pi^{(3)}(\omega-\omega_f-i0^+,\omega+\Omega+i0^+,\omega-i0^+)
    \\
    \nonumber
   & -
    \Pi^{(3)}(\omega-\omega_f-i0^+,\omega+\Omega-i0^+,\omega-i0^+)
    \\
    \nonumber
    &+
    \Pi^{(3)}(\omega+\omega_i-i0^+,\omega+\Omega+i0^+,\omega-i0^+)
    \\
    \nonumber
   & -
    \Pi^{(3)}(\omega+\omega_i-i0^+,\omega+\Omega-i0^+,\omega-i0^+)
    \\
    \nonumber
    &+
    \Pi^{(3)}(\omega+\omega_i+i0^+,\omega+\Omega+i0^+,\omega-i0^+)
    \\
    \nonumber
   & -\left.
    \Pi^{(3)}(\omega+\omega_i+i0^+,\omega+\Omega-i0^+,\omega-i0^+)
      \right\} .
\end{align}
Here the analytic continuation of Eq.~(\ref{Pi3A1g}) is
\begin{align}\label{Pi3A1g_c}
 &   \Pi^{(3)}(\omega_1,\omega_2,\omega_3)
    \\
    \nonumber
  &  =
    \frac{t^{*2}}{2(\omega_2-\omega_3)}
        \frac{\Sigma(\omega_3)-\Sigma(\omega_2)}{G(\omega_3)-G(\omega_2)}
        \\
        \nonumber
    &   \times\!
        \left[
        G(\omega_2)\frac{G(\omega_3)-G(\omega_1)}{Z(\omega_1)-Z(\omega_3)} -
        G(\omega_3)\frac{G(\omega_2)-G(\omega_1)}{Z(\omega_1)-Z(\omega_2)}
        \right].
\end{align}

\subsection{Resonant scattering}

\begin{figure}[!htbf]
    \begin{center}
        \includegraphics[width=0.68\columnwidth]{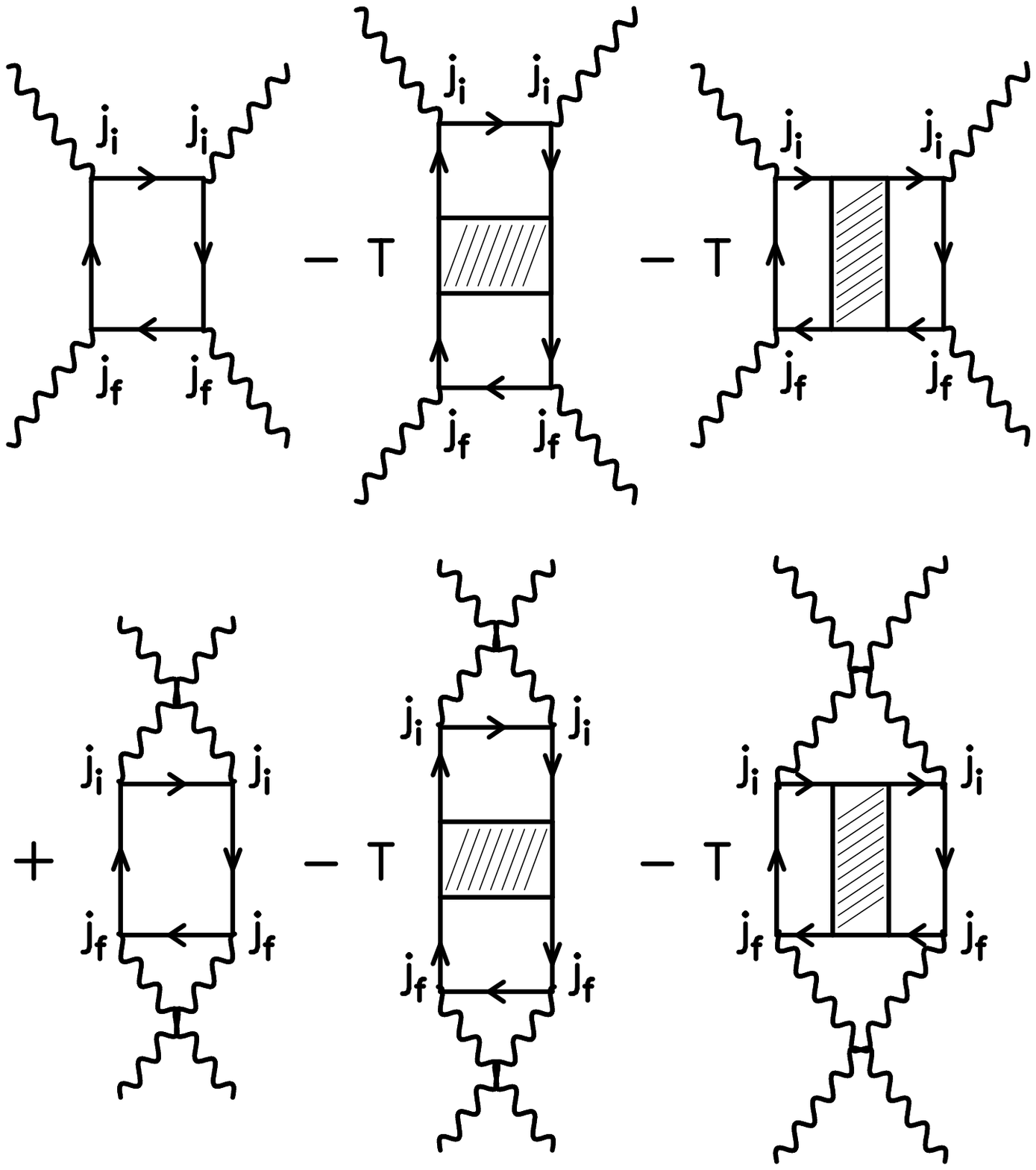}\\
        \includegraphics[width=0.68\columnwidth]{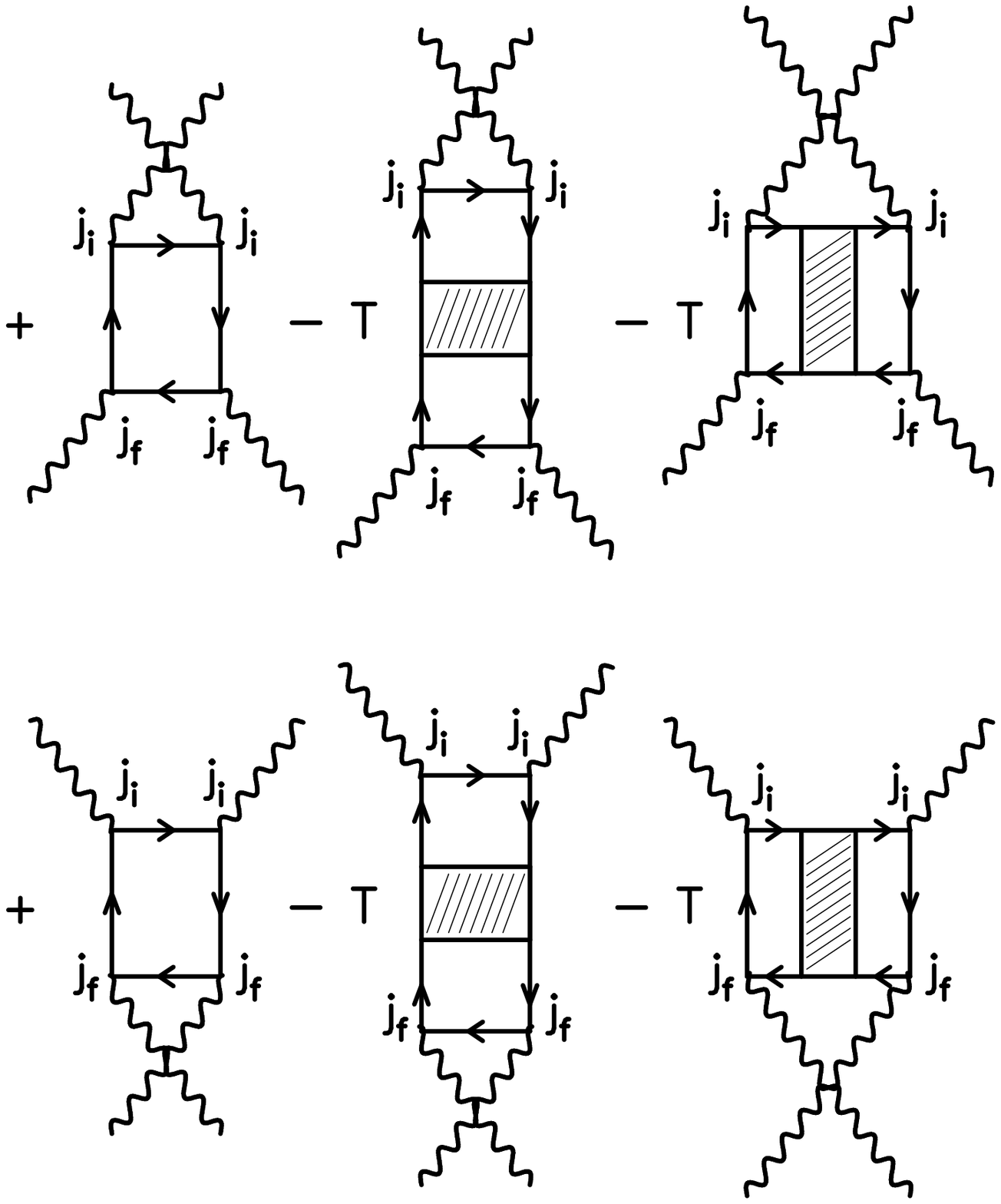}\\
        \includegraphics[width=0.73\columnwidth]{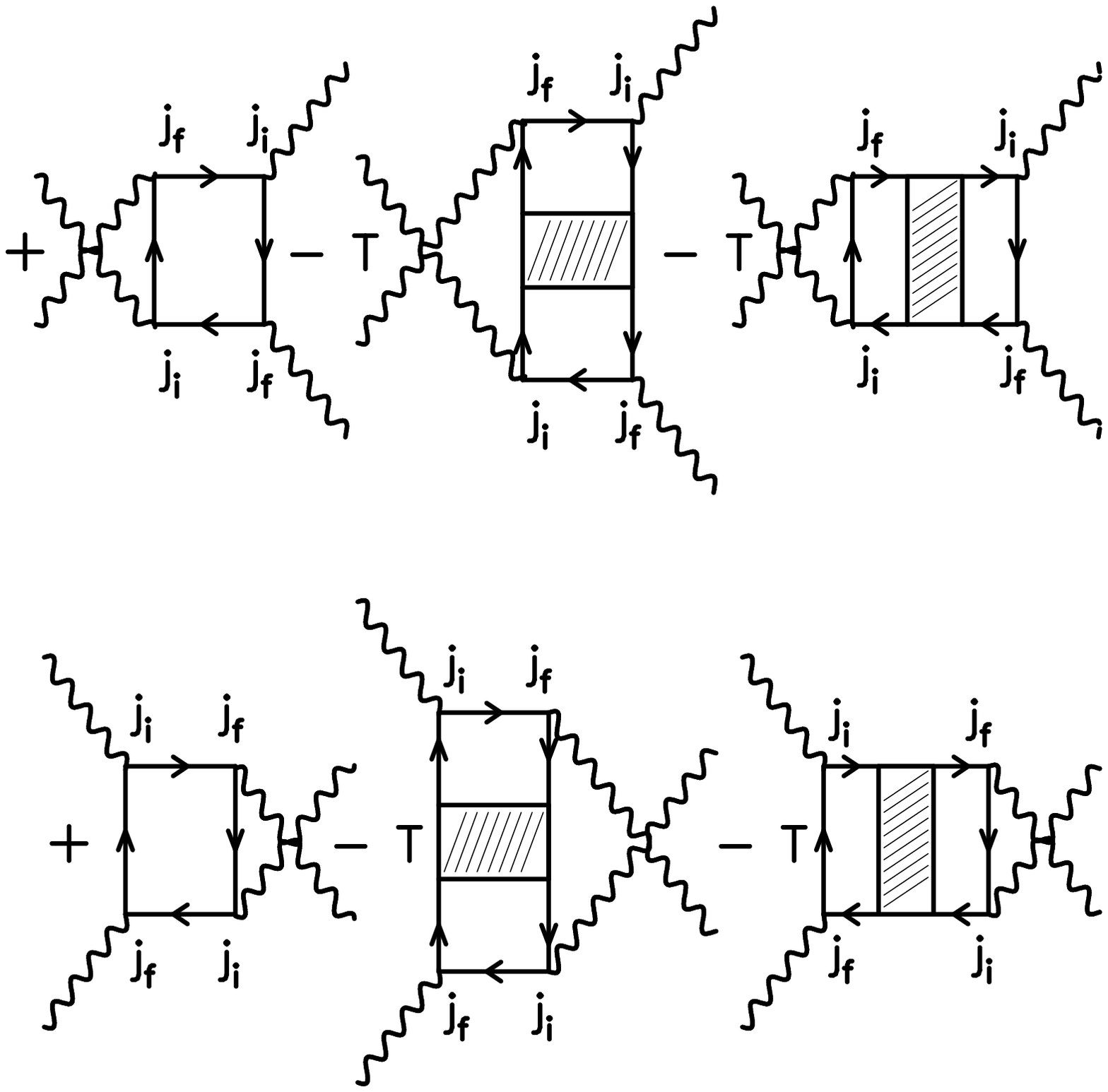}
    \end{center}
    \caption{Feynman diagrams for the resonant contributions to
Raman scattering. Only the first two diagrams in the first four lines, and
the first diagram in the last two lines contribute in the
$B_{\textrm{1g}}$ and $B_{\textrm{2g}}$ sectors. The $A_{\textrm{1g}}$
response includes all diagrams.}
    \label{fig:resonant_diagram}
\end{figure}

The resonant Raman response corresponds to scattering processes that
involve four external current vertices. The correlation function
constructed from four current operators contains six types of diagrams
corresponding to the different direct and exchange processes (see
Fig.~\ref{fig:resonant_diagram}). It should be noted that since $j^{(i)}$
and $j^{(f)}$ are odd functions of momentum, the only way to get a nonzero
momentum summation is to have an even number of current operators in any
given momentum integration (``current-operator pairing''). Hence all local
three-particle and four-particle vertex renormalizations must vanish,
although two-particle vertex renormalizations are possible. For the
$B_{1g}$ and $B_{2g}$ channels the ``current-operator pairing'' is
possible only between either both incoming $j^{(i)}$ or both final
$j^{(f)}$ current operators, but for the $A_{1g}$ channel all operators
can be involved in the ``pairing'' and the contribution from the bare
diagrams in the $A_{1g}$ channel is three times larger than for the
$B_{1g}$ channel. As a result, in the $B_{1g}$ and $B_{2g}$ channels we
have contributions from the first two diagrams in the first four lines and
from only the first diagram in the last two lines of
Fig.~\ref{fig:resonant_diagram}, and in the $A_{1g}$ channel all diagrams
contribute.

For the $B_{1g}$ and $B_{2g}$ channels, the product $j^{(i)}j^{(f)}$ is
orthogonal to the charge vertex with $A_{1g}$ symmetry, so the diagrams
are not renormalized across the vertices that contain both $j^{(i)}$ and
$j^{(f)}$ factors. In addition, for the $B_{2g}$ channel, the polarization
vectors select either odd or even momentum coordinates and, as a result,
the resonant Raman response for the $B_{2g}$ channel is four times smaller
than for the $B_{1g}$ one, and it is the only contribution to the total
Raman response in the $B_{2g}$ channel. In the $A_{1g}$ channel, besides
the diagrams presented in Fig.~\ref{fig:resonant_diagram} that include all
possible horizontal and vertical ``ladder'' renormalizations, one could
renormalize by parquet-like terms that involve simultaneous horizontal and
vertical renormalizations. But it can be shown (see the Appendix), that
such contributions are $1/D$ corrections, and disappear in the
$D\to\infty$ limit.

As a result, the Fourier transform of the four-time correlation function
constructed from the current operators can be represented in the following
form ($i\nu_i-i\nu_f=i\nu'_i -i\nu'_f $):
\begin{align} \label{chi4pi}
&   \chi^{(4)}_{i,f,f,i}(-i\nu_i, i\nu_f, -i\nu'_f, i\nu'_i) = T \sum_m
    \\
    \nonumber
    &\times\left[
    \Pi^{(4)}_I(i\omega_m,i\omega_m-i\nu_f,i\omega_m+i\nu_i-i\nu_f,i\omega_m-i\nu'_f)
    \right.
    \\
    \nonumber
    &+
    \Pi^{(4)}_I(i\omega_m,i\omega_m+i\nu'_f,i\omega_m-i\nu_i+i\nu_f,i\omega_m+i\nu_f)
    \\
    \nonumber
    &+
    \Pi^{(4)}_I(i\omega_m,i\omega_m-i\nu_f,i\omega_m-i\nu'_i-i\nu_f,i\omega_m-i\nu'_f)
    \\
    \nonumber
    &+
    \Pi^{(4)}_I(i\omega_m,i\omega_m+i\nu'_f,i\omega_m+i\nu_i+i\nu'_f,i\omega_m+i\nu_f)
    \\
    \nonumber
    &+
    \Pi^{(4)}_{II}(i\omega_m,i\omega_m+i\nu_i,i\omega_m+i\nu_i-i\nu_f,i\omega_m-i\nu'_f)
    \\
    \nonumber
    &+ \left.
    \Pi^{(4)}_{II}(i\omega_m,i\omega_m-i\nu_f,i\omega_m+i\nu_i-i\nu_f,i\omega_m+i\nu'_i)
    \right].
\end{align}

In the $B_{1g}$ and $B_{2g}$ channels, $\Pi^{(4)}_{II}$ contains only the
bare contribution (corresponding to only the first diagram on last two
lines of Fig.~\ref{fig:resonant_diagram})
\begin{align}\label{Pi4B1g}
        \Pi^{(4)}_{II,B_{1g}}(i\omega_1,i\omega_2,i\omega_3,i\omega_4) &=
        \Pi^{(4)}_{\text{bare}}(i\omega_1,i\omega_2,i\omega_3,i\omega_4),
        \\
        \nonumber
        \Pi^{(4)}_{II,B_{2g}}(i\omega_1,i\omega_2,i\omega_3,i\omega_4) &=
        \frac 14
        \Pi^{(4)}_{II,B_{1g}}(i\omega_1,i\omega_2,i\omega_3,i\omega_4) ,
\end{align}
where
\begin{align}
        \Pi^{(4)}_{\text{bare}}(i\omega_1,i\omega_2,&i\omega_3,i\omega_4)
    \\
    \nonumber
        =
        -\frac{t^{*4}}{D^2} \frac 1N & \sum_{\bm k}
        \frac{\sum\limits_{\alpha=1}^{D}\sin^2 k_\alpha \sum\limits_{\beta=1}^{D}\sin^2 k_\beta}
        {(Z_1-\epsilon_{\bm k})(Z_2-\epsilon_{\bm k})(Z_3-\epsilon_{\bm k})(Z_4-\epsilon_{\bm k})}
        \\
        \nonumber
        = - \frac{t^{*4}}{4} &\left[ \frac{G_1}{(Z_2-Z_1)(Z_3-Z_1)(Z_4-Z_1)}
        \right.
    \\
    \nonumber
        &+
        \frac{G_2}{(Z_1-Z_2)(Z_3-Z_2)(Z_4-Z_2)}
        \\
        \nonumber
        &+  \frac{G_3}{(Z_1-Z_3)(Z_2-Z_3)(Z_4-Z_3)}
    \\
    \nonumber
        &\left.
        + \frac{G_4}{(Z_1-Z_4)(Z_2-Z_4)(Z_3-Z_4)} \right ].
\end{align}
However, the other polarization $\Pi^{(4)}_I$ contains a vertical
``ladder'' renormalization (corresponding to the first two diagrams on
the first four lines of Fig.~\ref{fig:resonant_diagram})
\begin{align}
        \Pi^{(4)}_{I,B_{1g}}(i\omega_1,i\omega_2,i\omega_3,i\omega_4) &
    \\
    \nonumber
        =
        \Pi^{(4)}_{\text{bare}}(i\omega_1,i\omega_2,i\omega_3,i\omega_4)
        &+
        \Pi^{(4)}_{\text{r}}(i\omega_1,i\omega_2,i\omega_3,i\omega_4)
        ,
        \\
        \nonumber
        \Pi^{(4)}_{I,B_{2g}}(i\omega_1,i\omega_2,i\omega_3,i\omega_4) &=
        \frac 14
        \Pi^{(4)}_{I,B_{1g}}(i\omega_1,i\omega_2,i\omega_3,i\omega_4)
\end{align}
with
\begin{align}
        \Pi^{(4)}_{\text{r}}&(i\omega_1,i\omega_2,i\omega_3,i\omega_4)
    \\
    \nonumber
        =& - T
        \left(\frac1N \sum\limits_{\bm k} \frac{t^{*2}}{D} \sum\limits_{\alpha=1}^D
        \sin^2 k_\alpha G_1(\bm k) G_2(\bm k) G_4(\bm k)
        \right)
        \\
        \nonumber
        \times &
        \Tilde\Gamma(2, 4)
        \left(\frac1N \sum\limits_{\bm k} \frac{t^{*2}}{D} \sum\limits_{\alpha=1}^D
        \sin^2 k_\alpha G_2(\bm k) G_4(\bm k) G_3(\bm k)
        \right)
        \\
        \nonumber
        =& - T \frac{t^{*4}}{4}
        \left(
        \frac{G_4-G_1}{Z_1-Z_4} - \frac{G_2-G_1}{Z_1-Z_2}
        \right)
        \frac{\Tilde\Gamma(2, 4)}{(Z_2-Z_4)^2}
    \\
    \nonumber
        &\times\left(
        \frac{G_4-G_3}{Z_3-Z_4} - \frac{G_2-G_3}{Z_3-Z_2}
        \right).
\end{align}
Using the solution of Eq.~(\ref{gamma}) in the Bethe-Salpeter-like
equation (\ref{BSvrtx}) yields
\begin{align} \label{Pi4r}
        \Pi^{(4)}_{\text{r}}(i\omega_1,i\omega_2,i\omega_3,i\omega_4) &=
        \Tilde\Pi^{(4)}_{\text{r}}(i\omega_1,i\omega_2,i\omega_3,i\omega_4)
        \\
        \nonumber
        &
        + \frac{\Psi^{(4)}_{\text{r}}(i\omega_1,i\omega_2,i\omega_3,i\omega_4)}
        {i\omega_2-i\omega_4} ,
\end{align}
with
\begin{align}
    &   \Tilde\Pi^{(4)}_{\text{r}}(i\omega_1,i\omega_2,i\omega_3,i\omega_4) =
        \frac{t^{*4}}{4} \frac1{(G_2-G_4)(Z_2-Z_4)}
        \\
        \nonumber
    &   \times
        \left(
        \frac{G_4-G_1}{Z_1-Z_4} - \frac{G_2-G_1}{Z_1-Z_2}
        \right)
        \left(
        \frac{G_4-G_3}{Z_3-Z_4} - \frac{G_2-G_3}{Z_3-Z_2}
        \right),
\end{align}
and
\begin{align}
    &   \Psi^{(4)}_{\text{r}}(i\omega_1,i\omega_2,i\omega_3,i\omega_4) =
        -\frac{t^{*4}}{4} \frac1{G_2-G_4}
        \\
        \nonumber
    &   \times
        \left(
        \frac{G_4-G_1}{Z_1-Z_4} - \frac{G_2-G_1}{Z_1-Z_2}
        \right)
        \left(
        \frac{G_4-G_3}{Z_3-Z_4} - \frac{G_2-G_3}{Z_3-Z_2}
        \right)
        \\
        \nonumber
    &   =\Psi^{(4)}_{\text{r}}(i\omega_3,i\omega_2,i\omega_1,i\omega_4)
        =-\Psi^{(4)}_{\text{r}}(i\omega_1,i\omega_4,i\omega_3,i\omega_2) .
\end{align}

In the $A_{1g}$ channel we have contributions from all the diagrams in
Fig.~\ref{fig:resonant_diagram}, hence
\begin{align}\label{Pi4A1g}
    &   \Pi^{(4)}_{I,A_{1g}} (i\omega_1,i\omega_2,i\omega_3,i\omega_4) =
        \Pi^{(4)}_{II,A_{1g}} (i\omega_1,i\omega_2,i\omega_3,i\omega_4)
        \\
        \nonumber
        &=
        3 \Pi^{(4)}_{\text{bare}} (i\omega_1,i\omega_2,i\omega_3,i\omega_4) +
        \Pi^{(4)}_{\text{r}} (i\omega_1,i\omega_2,i\omega_3,i\omega_4)
        \\
        \nonumber
    &   +
        \Pi^{(4)}_{\text{r}} (i\omega_2,i\omega_3,i\omega_4,i\omega_1).
\end{align}
Here the last term corresponds to the horizontal ``ladder''
renormalization (the last diagram on each line of
Fig.~\ref{fig:resonant_diagram}).

Next, we perform the analytic continuation in Eq.~(\ref{chi4pi}) and
replace the sum over Matsubara frequencies by an integral over the real
axis in the same way as was done in Eq.~(\ref{chi3mr}) for the mixed
scattering. Then we substitute it into the expression in
Eq.~(\ref{chiRfin}) for the resonant Raman response. After some tedious
algebra, we achieve the final expression for the resonant Raman response
of the $D=\infty$ Falicov-Kimball model:
\begin{widetext}
\begin{align}\label{chiRfinFK}
    \chi_R(\Omega) =
    \frac2{(2\pi i)^2}& \int^{+\infty}_{-\infty} d\omega
    [f(\omega)-f(\omega+\Omega)]
    \\
    \nonumber
    \times\Real & \left\{
    \Pi^{(4)}_{I}(\omega-i0^+,\omega-\omega_f-i0^+,\omega+\Omega+i0^+,\omega-\omega_f+i0^+)
    \right.
    \\
    \nonumber
    &- \Pi^{(4)}_{I}(\omega-i0^+,\omega-\omega_f-i0^+,\omega+\Omega-i0^+,\omega-\omega_f+i0^+)
    \\
    \nonumber
    &+
    \Pi^{(4)}_{I}(\omega-i0^+,\omega+\omega_i-i0^+,\omega+\Omega+i0^+,\omega+\omega_i+i0^+)
    \\
    \nonumber
    &- \Pi^{(4)}_{I}(\omega-i0^+,\omega+\omega_i-i0^+,\omega+\Omega-i0^+,\omega+\omega_i+i0^+)
    \\
    \nonumber
    &+
    \Pi^{(4)}_{II}(\omega-i0^+,\omega+\omega_i+i0^+,\omega+\Omega+i0^+,\omega-\omega_f+i0^+)
    \\
    \nonumber
    &- \Pi^{(4)}_{II}(\omega-i0^+,\omega+\omega_i+i0^+,\omega+\Omega-i0^+,\omega-\omega_f+i0^+)
    \\
    \nonumber
    &+
    \Pi^{(4)}_{II}(\omega-i0^+,\omega+\omega_i-i0^+,\omega+\Omega+i0^+,\omega-\omega_f-i0^+)
    \\
    \nonumber
    &-\left.
    \Pi^{(4)}_{II}(\omega-i0^+,\omega+\omega_i-i0^+,\omega+\Omega-i0^+,\omega-\omega_f-i0^+)
      \right\} .
\end{align}
The analytic continuation in Eq.~(\ref{chiRfinFK}) can be found simply by
substituting $i\omega_\alpha\to\omega_\alpha\pm i0^+$ in the corresponding
expressions in Eqs.~(\ref{Pi4B1g})--(\ref{Pi4A1g}) which will not be
explicitly repeated here. It might appear that the first four terms in
braces contain divergences connected with vanishing denominators in the
last term in Eq.~(\ref{Pi4r}), but the contribution of these terms into
the expression in braces in Eq.~(\ref{chiRfinFK}) must be considered in
the limit:
\begin{align}\label{psiRlim}
    \lim_{\Delta\to0} \frac1{2\Delta} &\left\{
    \Psi^{(4)}_{\text{r}}
    (\omega-i0^+,\omega-\omega_f-i0^+,\omega+\Omega+i0^+,\omega-\omega_f-\Delta+i0^+)
    \right.
    \\
    \nonumber
    &-
    \Psi^{(4)}_{\text{r}}
    (\omega+i0^+,\omega-\omega_f-i0^+,\omega+\Omega+i0^+,\omega-\omega_f-\Delta+i0^+)
    \\
    \nonumber
    &-
    \Psi^{(4)}_{\text{r}}
    (\omega-i0^+,\omega-\omega_f-i0^+,\omega+\Omega-i0^+,\omega-\omega_f-\Delta+i0^+)
    \\
    \nonumber
    &+
    \Psi^{(4)}_{\text{r}}
    (\omega+i0^+,\omega-\omega_f-i0^+,\omega+\Omega-i0^+,\omega-\omega_f-\Delta+i0^+)
    \\
    \nonumber
    &+
    \Psi^{(4)}_{\text{r}}
    (\omega-i0^+,\omega+\omega_i+\Delta-i0^+,\omega+\Omega+i0^+,\omega+\omega_i+i0^+)
    \\
    \nonumber
    &-
    \Psi^{(4)}_{\text{r}}
    (\omega+i0^+,\omega+\omega_i+\Delta-i0^+,\omega+\Omega+i0^+,\omega+\omega_i+i0^+)
    \\
    \nonumber
    &-
    \Psi^{(4)}_{\text{r}}
    (\omega-i0^+,\omega+\omega_i+\Delta-i0^+,\omega+\Omega-i0^+,\omega+\omega_i+i0^+)
    \\
    \nonumber
    &+\left.
    \Psi^{(4)}_{\text{r}}
    (\omega+i0^+,\omega+\omega_i+\Delta-i0^+,\omega+\Omega-i0^+,\omega+\omega_i+i0^+)
      \right\} ,
\end{align}
where $\Delta=\omega'_f-\omega_f=\omega'_i-\omega_i$. When the limit
$\Delta\to0$ is taken, we find that the imaginary part of
Eq.~(\ref{psiRlim}) diverges, but the real part (which is all that
contributes to the Raman scattering) is finite and can be calculated using
l'Hopital's rule:
\begin{align}\label{psiRdrv}
    -\frac12 \Real &\left\{
    \Tilde\Psi'
    (\omega-i0^+,\omega-\omega_f-i0^+,\omega+\Omega+i0^+,\omega-\omega_f+i0^+)
    \right.
    \\
    \nonumber
    &-
    \Tilde\Psi'
    (\omega-i0^+,\omega-\omega_f+i0^+,\omega+\Omega-i0^+,\omega-\omega_f-i0^+)
    \\
    \nonumber
    &-
    \Tilde\Psi'
    (\omega-i0^+,\omega-\omega_f-i0^+,\omega+\Omega-i0^+,\omega-\omega_f+i0^+)
    \\
    \nonumber
    &+
    \Tilde\Psi'
    (\omega-i0^+,\omega-\omega_f+i0^+,\omega+\Omega+i0^+,\omega-\omega_f-i0^+)
    \\
    \nonumber
    &+
    \Tilde\Psi'
    (\omega-i0^+,\omega+\omega_i+i0^+,\omega+\Omega+i0^+,\omega+\omega_i-i0^+)
    \\
    \nonumber
    &-
    \Tilde\Psi'
    (\omega-i0^+,\omega+\omega_i-i0^+,\omega+\Omega-i0^+,\omega+\omega_i+i0^+)
    \\
    \nonumber
    &-
    \Tilde\Psi'
    (\omega-i0^+,\omega+\omega_i+i0^+,\omega+\Omega-i0^+,\omega+\omega_i-i0^+)
    \\
    \nonumber
    &+\left.
    \Tilde\Psi'
    (\omega-i0^+,\omega+\omega_i-i0^+,\omega+\Omega+i0^+,\omega+\omega_i+i0^+)
      \right\} ,
\end{align}
\end{widetext}
with
\begin{equation}
        \Tilde\Psi'(1,2,3,4)=\frac{d\Psi^{(4)}_{\text{r}}(1,2,3,4)}{dZ_4}
        \frac{dZ_4}{d\omega_4}
\end{equation}
and
\begin{align}
        \frac{d\Psi^{(4)}_{\text{r}}(1,2,3,4)}{dZ_4} =
        -\frac{t^{*4}}{4}
        \frac{dG_4/dZ_4}{(G_2-G_4)^2}
        \\
        \nonumber
        \times
        \left(
        \frac{G_4-G_1}{Z_1-Z_4} - \frac{G_2-G_1}{Z_1-Z_2}
        \right)
        \left(
        \frac{G_4-G_3}{Z_3-Z_4} - \frac{G_2-G_3}{Z_3-Z_2}
        \right)
        \\
        \nonumber
        -
        \frac{t^{*4}}{4}
        \frac1{G_2-G_4}
        \left(
        \frac{dG_4/dZ_4}{Z_1-Z_4} + \frac{G_4-G_1}{(Z_1-Z_4)^2}\right)
        \\
        \nonumber
        \times
        \left(
        \frac{G_4-G_3}{Z_3-Z_4} - \frac{G_2-G_3}{Z_3-Z_2}
        \right)
        \\
        \nonumber
        -
        \frac{t^{*4}}{4}
        \frac1{G_2-G_4}
        \left(
        \frac{G_4-G_1}{Z_1-Z_4} - \frac{G_2-G_1}{Z_1-Z_2}
        \right)
        \\
        \nonumber
        \times
        \left(
        \frac{dG_4/dZ_4}{Z_3-Z_4} + \frac{G_4-G_3}{(Z_3-Z_4)^2}
        \right),
\end{align}
\begin{equation}
        \frac{dG}{dZ} = \frac2{t^{*2}}(1-ZG),
\end{equation}
\begin{equation}
        \frac{dZ}{d\omega} = 1 - \frac{d\Sigma}{d\omega}.
\end{equation}
When the ground state of the ``paramagnetic'' phase is metallic, the
derivative of the self energy is straightforward to calculate as
\begin{align}
        \frac{d\Sigma}{d\omega} & = \frac{2}{t^{*2}}\;\frac{1-ZG}{G^2}
        \\
        \nonumber
    &   \times
        \frac{(\Sigma-U w_1)^2}
        {U^2w_1(1-w_1)+(\Sigma-U w_1)^2(1+\frac{2}{t^{*2}}\frac{1-ZG}{G^2})}.
\end{align}
but in the ``insulating'' phase, when $U^2 w_1 (1-w_1) > t^{*2}/2$ at any
filling $w_1$, one has to include the contribution from the delta-function
peak in the imaginary part of the self-energy which yields the additional
contribution
\begin{align}
\label{sigma_deriv_ins}
        \frac{d\Sigma}{d\omega} & = \frac{2}{t^{*2}}\;\frac{1-ZG}{G^2}
        \\
        \nonumber
    &   \times
        \frac{(\Sigma-U w_1)^2}
        {U^2w_1(1-w_1)+(\Sigma-U w_1)^2(1+\frac{2}{t^{*2}}\frac{1-ZG}{G^2})}
        \\
        \nonumber
    &   \pm i\pi \left[ U^2 w_1 (1-w_1) - t^{*2}/2 \right]\frac{d\delta(\omega-U[1-w_1])}{d\omega},
\end{align}
to the derivative.

\subsection{Bare contributions and multiple resonances}

In summary, the total Raman response function is the sum of the nonresonant
[Eq.~(\ref{RNfinFK})], mixed [Eq.~(\ref{RMfinFK})], and resonant 
[Eq.~(\ref{chiRfinFK})]
contributions and has a complicated form. It is educational to consider the
contributions of the bare
diagrams, which can be summed up and rewritten in the following 
form\cite{shvaika_conf1}:
\begin{align}\label{chi_bare}
    \chi(\Omega)=\frac1N\sum_{\bm k}\int^{+\infty}_{-\infty} d\omega
    [f(\omega)-f(\omega+\Omega)] A_{\bm k}(\omega) A_{\bm k}(\omega+\Omega)
    \\
    \nonumber
    \times\left|\gamma_{\bm k}+v_{\bm k}^{i}v_{\bm k}^{f}
    \left[G_{\bm k}(\omega+\omega_i+i0^+)+G_{\bm k}(\omega-\omega_f-i0^+)\right]\right|^2,
\end{align}
where $\gamma_{\bm k}=\sum_{\alpha,\beta} e_\alpha^i
\frac{\partial^2\epsilon_{\bm k}}{\partial k_\alpha\partial k_\beta}
e_\beta^f$, $v_{\bm k}^{i,f}=\sum_{\alpha} e_\alpha^{i,f}
\frac{\partial\epsilon_{\bm k}}{\partial k_\alpha}$, $A_{\bm
k}(\omega)=\frac1\pi\Img G_{\bm k}(\omega-i0^+)$, and
\begin{equation}
    G_{\bm k}(\omega)=\frac1{\omega+\mu-\Sigma(\omega)-\epsilon_{\bm k}}
\end{equation}
is the momentum-dependent Green's function.

In general, the bare response function in Eq.~(\ref{chi_bare}) is a function of 
the frequency shift $\Omega=\omega_i-\omega_f$, of the incoming photon 
frequency $\omega_i$ and the outgoing photon $\omega_f$ frequency; it
can be enhanced when one or both of the denominators are resonant (i.e., they
coincide). In the
latter case, we have a so-called ``double'' or ``multiple
resonance''.\cite{MartinFalicov} The full response function also includes the
vertex renormalizations. But the total (reducible) charge vertex in 
Eq.~(\ref{gamma}) for the Falicov-Kimball model does not diverge, and hence it 
does not introduce any
additional ``resonances.'' It only leads to a renormalization of the
total Raman response.

\section{Numerical Results}

\begin{figure}[htbf]
\includegraphics[width=0.80\columnwidth]{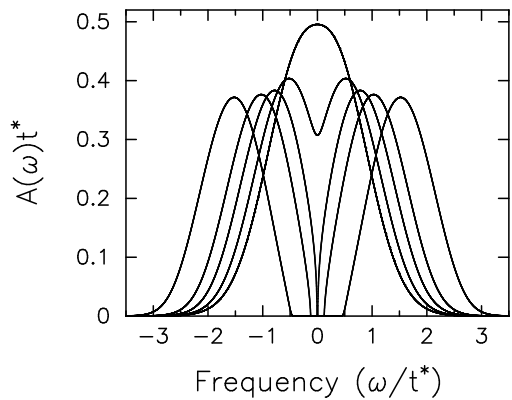}
\caption{\label{fig: dos} Interacting single-particle density of states
for $U=0.5$, 1.0, 1.5, 2.0, and 3.0 ($U$ increases as the pseudogap
becomes stronger). Note how the DOS first develops a depression near the
chemical potential and then develops a pseudogap as the metal-insulator
transition occurs (the DOS vanishes only at $\omega=0$ in the insulator).}
\end{figure}

We begin our results by showing the single particle density of states of
the spinless Falicov-Kimball model in infinite dimensions with $\langle
\rho_e\rangle=\langle w_i\rangle =1/2$.  The density of states is
independent of temperature, and a metal-insulator transition occurs at
$U=\sqrt{2}$.  In the insulating phase, the self energy develops a pole at
$\omega=0$, and the Green's function vanishes there.  There is no true gap
to this system, as the bare Gaussian density of states forces the
interacting density of states to be nonzero whenever the self energy is
finite.\cite{prl_mit}
In Fig.~\ref{fig: dos}, we plot the DOS for 5 values of $U$
ranging from a weakly scattering metal $U=0.5$, to a strongly scattering
metal $U=1$, to a near-critical insulator $U=1.5$, a ``small-gap''-insulator
$U=2$ and a ``moderate-gap''-insulator $U=3$.  Note that the metal-insulator
transition is continuous for the Falicov-Kimball model, in the sense that
the zero-temperature dc conductivity continuously goes to zero at the
transition.  Note further that in the metallic phase, the system is not a
Fermi liquid because the scattering time at the putative Fermi surface
does not become infinite as $T\rightarrow 0$.

Once the self energy and the DOS are known, the different contributions to
Raman scattering can be determined by straightforward, but tedious
numerical integrations of the relevant functions for each scattering
channel [Eqs.~(\ref{RNfinFKB1g}, \ref{RNfinFKA1g}, \ref{RMfinFK},
\ref{Pi3A1g_c}, and \ref{chiRfinFK}--\ref{sigma_deriv_ins}). 
There are some subtleties with this approach, especially in the
insulating phase, as the iterative approach to determining the DOS and the
self energy becomes inaccurate once the imaginary part of the self energy
becomes smaller than about $10^{-13}$.  Fortunately, there is a simple
analytic form that can be used to construct the imaginary parts of the
Green's functions and self energies in this regime, so all relevant
quantities can be evaluated with care.\cite{prb_thermal}

We find that the Stokes response is significantly larger than the
anti-Stokes response in the resonant regime, because the double resonance
greatly enhances the signal when the transfered energy approaches the
incident photon frequency (in the nonresonant regime, both Stokes and
anti-Stokes responses are identical). Hence, we will present only the
Stokes response here.  We also find that, generically, the response
``sharpens'' as $T\rightarrow 0$, with the spectral response growing at
low temperature (except for the low-energy, thermally excited response in
the insulating phase).  Thus, we focus on low and moderate temperatures in
the metallic regime, since the Raman signal is largest there.

\begin{figure}[htbf]
\includegraphics[width=0.80\columnwidth]{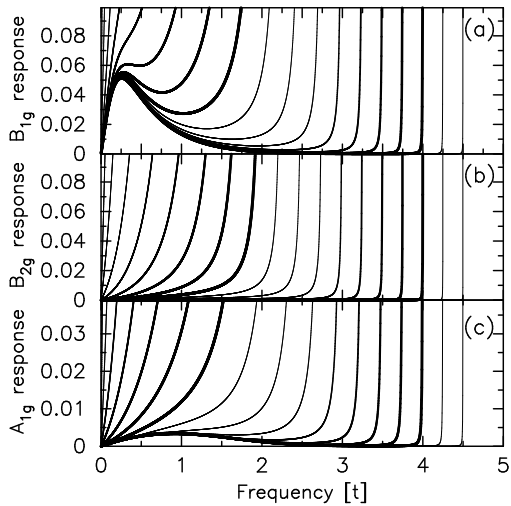}
\caption{\label{fig: channel_u=0.5} Stokes Raman response for the three
symmetry channels in a dirty metal with $U=0.5$.  The Raman scattering
response function is plotted as a function of the transfered frequency for
incident photon frequencies ranging from 0.25 to 4.5 in steps of 0.25 (the
thickness of the lines aids in distinguishing the different curves).  This
data is at low temperature ($T=0.05$) where the results are the
``sharpest''.}
\end{figure}

\begin{figure}[htbf]
\includegraphics[width=0.80\columnwidth]{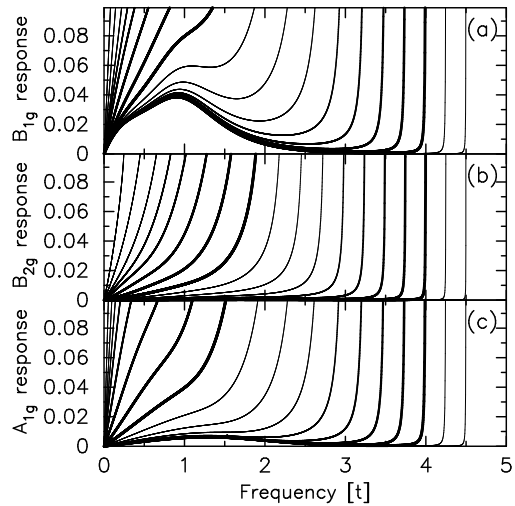}
\caption{\label{fig: channel_u=1.0} Stokes Raman response for the three
symmetry channels in a strongly scattering ``metal'' with $U=1.0$.  The
Raman scattering response function is plotted as a function of the
transfered frequency for incident photon frequencies ranging from 0.25 to
4.5 in steps of 0.25.  This data is at a moderate temperature ($T=0.5$)
where the nonresonant response has enhanced low-energy spectral weight in
the $B_{\textrm 1g}$ channel.}
\end{figure}

In Figs.~\ref{fig: channel_u=0.5} and \ref{fig: channel_u=1.0} we plot the
total Raman response for $U=0.5$ and $U=1$ respectively.  The former case
is of a dirty metal, while the latter case is a metal that has such strong
scattering that the density of states is depressed near the Fermi energy
(but not so much as to create an insulator).  The Stokes branch of the
Raman response behaves in many respects as expected.  The double resonance
causes a large enhancement of the signal as the transfered frequency
approaches the incident photon frequency. In the Loudon-Fleury regime,
where the photon energy is much larger than the band energies, one can see
a nice separation of the signal into the nonresonant and resonant (plus
mixed) pieces (note that the nonresonant $A_{\textrm 1g}$ response is
small due to screening effects and the nonresonant $B_{\textrm 2g}$
response vanishes due to symmetry, but the resonant effects are strong in
both of these channels). In general, the resonant effects are strongest
near the double resonance, and it is not true that the total response
looks like the nonresonant response plus a uniform resonant enhancement,
so resonant effects must be studied with care to understand the effects
they play on the light scattering. Finally, note the overall similarity
between panels (b) and (c) in Figs.~\ref{fig: channel_u=0.5} and \ref{fig:
channel_u=1.0}.  This arises from the fact that generically, the resonant
effects overwhelm both nonresonant effects and mixed scattering effects,
and it shows that there is not a huge variation in the resonant Raman
response due to the additional renormalizations in the $A_{\textrm 1g}$
channel.

\begin{figure}[htbf]
\includegraphics[width=0.80\columnwidth]{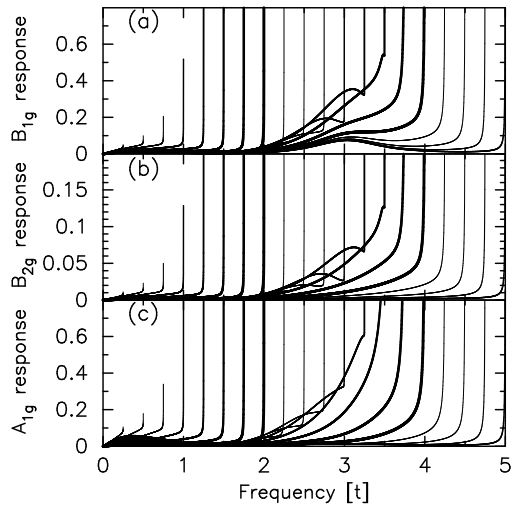}
\caption{\label{fig: channel_u=3.0} Stokes Raman response for the three
symmetry channels in a correlated insulator with $U=3.0$.  The Raman
scattering response function is plotted as a function of the transfered
frequency for incident photon frequencies ranging from 0.25 to 5.0 in
steps of 0.25.  This data is at a high temperature ($T=1.0$) where the
nonresonant response has enhanced low-energy spectral weight in
$B_{\textrm 1g}$ and $A_{\textrm 1g}$ channels.}
\end{figure}

\begin{figure}[htbf]
\includegraphics[width=0.80\columnwidth]{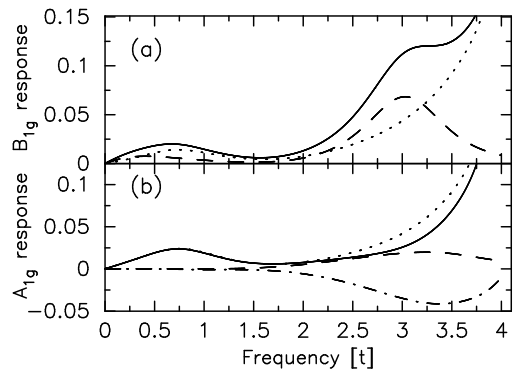}
\caption{\label{fig: separate} Separation of different contributions to
the Stokes response for $U=3$, $T=1.0$, and $\omega_i=4.0$ in the (a)
$B_{\textrm 1g}$ and (b) $A_{\textrm 1g}$ channels. The solid line is the
total response, the dotted line is the resonant piece, the dashed line is
the nonresonant piece, and the chain-dotted line is the mixed
contribution.}
\end{figure}

The insulating phase $(U>\sqrt{2})$ provides a number of interesting new
features to the electronic Raman scattering (results for the near-critical
insulator\cite{shvaika_conf2} at $U=1.5$ and for the small-gap 
insulator\cite{ll8979} at $U=2$ have already appeared).  We begin with 
a discussion
of a good correlated insulator $U=3$, which appears to have a well defined
gap region in the DOS (but note that the DOS only vanishes exactly at
$\omega=0$). Hence we expect there to be significant thermally driven
effects in this case.  To begin, we plot the Raman scattering at a fixed
temperature, but with varying incident photon frequency in Fig.~\ref{fig:
channel_u=3.0}. Note that there is substantial spectral weight in both a
low-energy and a high-energy peak, and that when the incident photon
frequency is approximately equal to $U$, the high-energy (charge-transfer)
peak can be enhanced significantly. But something strange occurs for
higher frequencies in the $A_{\textrm 1g}$ channel.  As $\omega_i$
increases beyond about 3.25, we stop to see the development of a separate
charge transfer peak, and the net scattering curve looks like a simple
double resonance curve even though the nonresonant response has a
well-developed charge transfer peak.  In other words, we are not seeing
the evolution of the scattering to a simple break up of a nonresonant
piece and a double resonance piece as $\omega_i$ is made large. This may
not be too surprising, because in the $A_{\textrm 1g}$ channel we have
nonresonant, resonant, and mixed contributions to the scattering. To
illustrate how this occurs, we plot the separate contributions to the
Raman scattering in Fig.~\ref{fig: separate} for the $B_{\textrm 1g}$ and
$A_\textrm{1g}$ channels for $\omega_i=4.0$.  In the top panel, we see the
expected shape for the nonresonant curve, with both low and high energy
peaks, but surprisingly, there is a strong resonant enhancement of both
peaks.  This is even more dramatic in the bottom panel, where the vertex
corrections suppress the nonresonant low-energy peak in the $A_{\textrm
1g}$ channel, but the resonant terms bring back a strong enhancement in
that region (in essence because the conservation of total charge acts to
effectively screen the low-energy excitations, but the screening is much
less effective for the resonant terms).  The mixed contribution is small
at low energy, but has a well developed charge-transfer-like feature, that
is negative, and completely overwhelms, and cancels the nonresonant
charge-transfer peak, leaving behind essentially a double resonance-like
curve. These results are obviously quite complex.  If the incident photon
frequency increases further, then the peak in the mixed response moves to
higher energy, and the nonresonant peak plus a higher frequency double
resonance peak picture holds, but the width of the double resonance peak
can be extremely narrow. One might be surprised that the double resonance
peak survives in the insulator (because there are no electronic states
within the gap), but in this case, we only have a pseudogap, and the
states in the ``gap region'' are few in number, but long-lived and hence
contribute to the scattering.\cite{prb_thermal}

\begin{figure}[htbf]
\includegraphics[width=0.80\columnwidth]{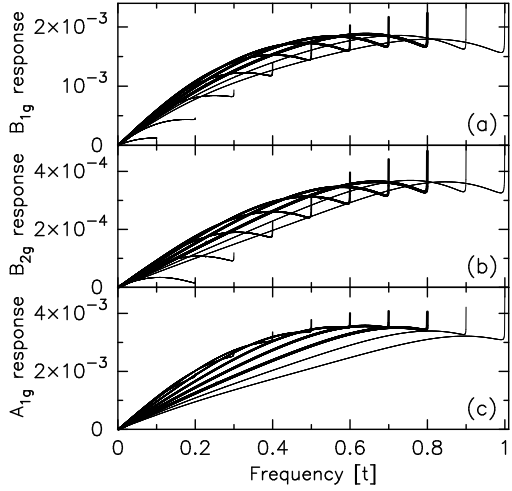}
\caption{\label{fig: channel_low} Raman response at low energy for $U=3$
and $T=0.2$.  The incident photon frequency changes from 0.1 to 1.0 in
steps of 0.1.}
\end{figure}

One of the common features in resonant Raman scattering is a large
enhancement of the scattering when a new scattering channel opens, as the
photon frequency becomes larger than an energy gap, for example.  One
question to ask is does such a feature survive in a correlated system.  As
described above, there is no energy gap in the insulating phase (on 
the hypercubic lattice), but there
is a region where the DOS is exponentially small, and then increases
rapidly to be of order unity.  One can ask whether there are features in
the Raman scattering that show enhancements when the photon frequency is
larger than the width of the exponentially small ``gap region'' of the
DOS.  Since the gap region for $U=3$ is about 0.5 above and below the
chemical potential, we expect interesting results for photon frequencies
near 0.5.  We plot the Raman scattering for $\omega_i$ increasing from 0.1
to 1 in steps of 0.1 in Fig.~\ref{fig: channel_low} for low temperature
($T=0.2$).  Note how small the overall scale of the Raman scattering is.
We see different behavior in the $B_{\textrm{1g}}$ and $B_{\textrm{2g}}$
sectors versus the $A_{\textrm{1g}}$ sector.  In panels (a) and (b) we see
the low energy scattering increases as $\omega_i$ increases until
$\omega_i$ reaches approximately 0.5, where it starts to decrease.  The
increasing behavior is essentially this resonant enhancement due to the
opening of scattering channels as the photon frequency becomes larger than
the gap. Note how this phenomenon essentially does not occur in panel (c),
where the curves lie below each other as $\omega_i$ is increased. Hence
the $A_{\textrm{1g}}$ channel does not show the analogue of this 
resonant-enhancement effect.  The effect disappears in all channels once the
temperature becomes larger than about 0.5, where thermal excitations can
be easily made across the ``gap region''. Note, furthermore, that the
largest resonant effects occur not when the scattering channel first
opens, but rather when $\omega_i\approx U$ because that is the value of
frequency that separates the peaks in the single-particle DOS, and hence
it corresponds to the strongest scattering from occupied to unoccupied
states.

\begin{figure}[htbf]
\includegraphics[width=0.80\columnwidth]{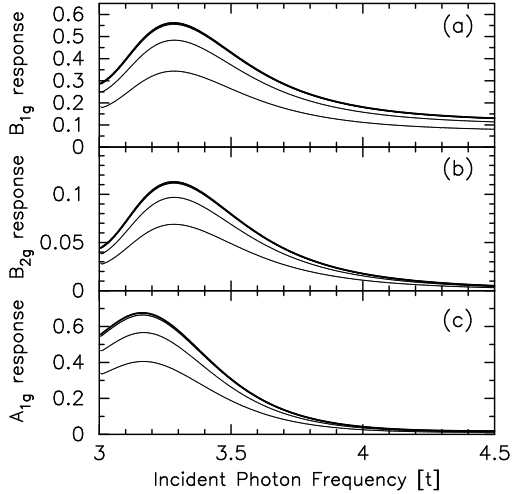}
\caption{\label{fig: profile_high} Raman response at $\Omega=3$ for $U=3$
and various temperatures.  The horizontal axis is the incident photon
frequency. The thickest curve is $T=0.05$, and the temperature increases
to $0.2$, $0.5$, and $1$ as the curves are made thinner. Note that the
curves for the lowest two temperatures are the largest, and are hard to
separate.}
\end{figure}

We saw in Fig.~\ref{fig: separate} that there is a resonant enhancement at
low energy when the incident photon frequency is close to $U$ in size. To
examine this phenomenon further, we plot the total Raman scattering at a
fixed transfered photon frequency (chosen to be $0.5$ for the low-energy
peak and $3.0$ for the high-energy peak) as a function of the incident
photon frequency in Figs.~\ref{fig: profile_high} and \ref{fig:
profile_low}.

\begin{figure}[htbf]
\includegraphics[width=0.80\columnwidth]{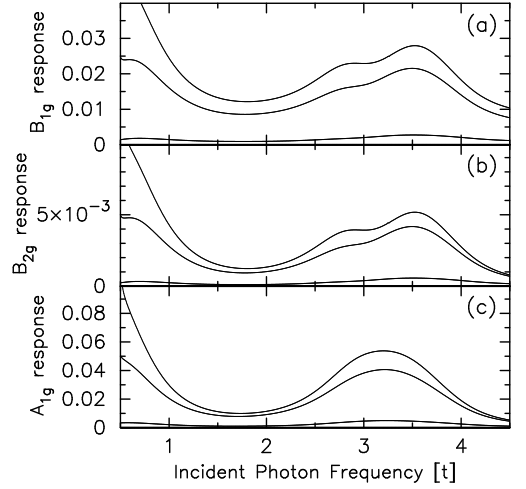}
\caption{\label{fig: profile_low} Raman response at $\Omega=0.5$ for $U=3$
and various temperatures.  The horizontal axis is the incident photon
frequency. The thickest curve is $T=0.05$, and the temperature increases
to $0.2$, $0.5$, and $1$ as the curves are made thinner. Note that the
$T=0.05$ curves are more than three orders of magnitude smaller than the
$T=0.2$ curves, and cannot be distinguished from the horizontal axis in
the figure.}
\end{figure}

In Fig.~\ref{fig: profile_high}, we see expected behavior.  The
charge-transfer peak at $\Omega=U=3$ has a resonant enhancement for photon
frequencies slightly higher than $U$, then a suppression to the
nonresonant peak values at the highest incident frequencies (except for
the $A_{\textrm{1g}}$ channel, where the charge transfer peak is initially
suppressed until the incident photon frequency is larger than about 6, due
to the cancellation from the mixed diagrams described above). The width of
the resonant peak is about 0.5, and it is pushed to higher frequency in
the $B_{\textrm{1g}}$ and $B_{\textrm{2g}}$ channels.  In Fig.~\ref{fig:
profile_low}, we find an interesting joint resonance effect. There is a
resonant enhancement near $\omega_i=0.5$, that comes from the double
resonance. In addition, there is another broad resonance effect centered
just slightly higher than $\omega_i=U=3$, where both the charge-transfer
and the low-energy peaks resonate at the same incident photon frequency.
In the $A_{\textrm{1g}}$ channel, the joint resonance peak is a single
smooth peak, while in the $B_{\textrm{1g}}$ and $B_{\textrm{2g}}$
channels, the joint resonance peak seems to have a double-peak structure
to it. As the temperature is reduced, the resonant effects remain, but the
spectral weight in the low-energy peak gets suppressed to very small
values (the $T=0.05$ curves are indistinguishable from the horizontal axis
because they are at least three orders of magnitude smaller than the
$T=0.2$ curves). The evolution of the resonant profile for other values
of transferred frequency $\Omega$ is complex and can be found 
in Ref.~\onlinecite{shvaika_conf1} for $U=3$.

\begin{figure}[htbf]
\includegraphics[width=0.80\columnwidth]{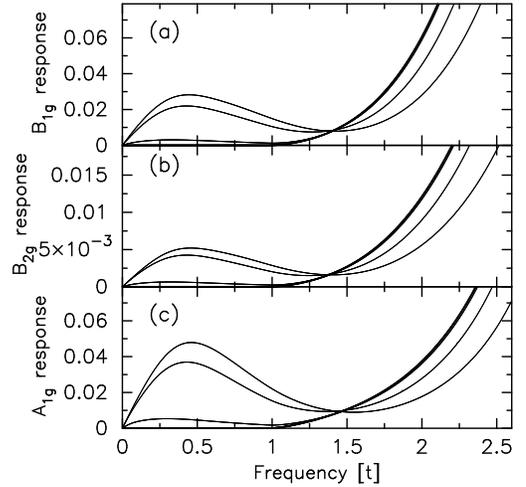}
\caption{\label{fig: iso_u=3} Low-energy 
isosbestic behavior of the resonant Raman
response for $U=3$ and $\omega_i=3.5$. The different thicknesses
correspond to different temperatures, with thicker curves corresponding to
lower temperatures (four curves are plotted for $T=1$, 0.5, 0.2, and
0.05). Note how the curves all cross at an isosbestic point, just slightly
smaller than $U/2$. }
\end{figure}

The low-energy isosbestic behavior (which means that the Raman response is
independent of temperature at a characteristic frequency) is plotted in
Fig.~\ref{fig: iso_u=3}. We choose $\omega_i=3.5$ because it corresponds
to the maximal joint resonance effect for both the charge-transfer and
low-energy peaks.  We find that the low-energy isosbestic behavior is generic 
for the resonant Raman scattering, with the response curves crossing at
$\Omega\approx U/2=1.5$ for all symmetries. Hence, the low-energy
isosbestic behavior
seen in the nonresonant response (which was most apparent in the
$B_{\textrm{1g}}$ channel, but can also be seen in the $A_{\textrm{1g}}$
channel when the response is plotted on a logarithmic 
scale\cite{shvaika_conf1}), survives in
the resonant cases as well, and this helps explain why it is seen in so
many experimental systems. In addition to the low-energy isosbestic point
shown in Fig.~\ref{fig: iso_u=3} at $\Omega\approx U/2$, there is a 
second isosbestic point\cite{shvaika_conf1} that appear near the double
resonance $\Omega\approx\omega_i$.  Starting from large $\omega_i$
($\Omega_i>U$), as $\omega_i$ is reduced, the two isosbestic points
move closer to each other, eventually joining together and disappearing
when $\omega_i\approx U/2$.  So the isosbestic behavior will not be
seen if the incident photon frequency is too low.

\section{Discussion}

With the use of DMFT, we solved for the full Raman response for all
frequencies of incoming light in the Falicov-Kimball model. Since the
Falicov-Kimball model can be tuned across a metal-insulator transition, we
have determined the form of Raman scattering in both the metallic and
insulating states, and have investigated light scattering on both sides of
the quantum critical point at $U=\sqrt{2}$. Resonant, non-resonant, and
mixed contributions have all been treated on an equal footing and we
allowed for an analysis of the dependence of Raman scattering with
temperature, interactions, and different light polarizations.

Our results confirm a number of previously held beliefs. First, we find a
strong resonant enhancement of the charge-transfer peak in Raman
scattering when the incident photon energy lies near the charge-transfer
energy. This behavior is robust to temperature and polarization changes
due to the local nature of the charge-transfer excitation in our model.
Second, we also find a polarization-independent ``double-resonance''
enhancement when the transfered frequency of the light approaches the
incident light frequency.  This feature survives in the insulating phase
because of the pseudogap nature of the insulator on the hypercubic lattice.

In addition, we find a number of new features of light scattering in
correlated insulators. We find that low energy spectral features, related
to thermal populations of elementary excitations, show resonance behavior
when the incident light is tuned to the much higher frequency of the
charge-transfer energy. This is a specific case where the correlations are
crucial, since in uncorrelated materials, this would correspond to
off-resonant conditions. Yet due to the many-body nature of the correlated
band, spectral features well separated from the charge transfer peak have
a non-trivial resonance profile. We believe that these may be potentially
useful to understand the complex nature of charge excitations in
correlated materials as it would impact both electronic and phononic Raman
scattering at low frequencies. Finally, we find that the presence of an
isosbestic point in the Raman response for correlated insulators results
from a symmetry-dependent combination of all resonant, mixed, and
non-resonant terms, and appears to be generic.

We close with a discussion of open questions concerning improvements to
the theory. Here we have restricted ourselves to Raman scattering in a
correlated band of electrons in the limit of large spatial dimensions.
Performing calculations in physical dimensions requires more many-particle
charge vertex renormalizations which makes the problem extremely
difficult, though possible in principle. But we found that most vertex
renormalizations were rather mild, so including nonlocal effects into the
vertices (finite dimensions) probably does not change these results
dramatically (unless the vertex can diverge in finite 
dimensions). In addition, $\bm q$-dependent information would prove to be
useful for investigating dispersive many-particle excitations, as probed
in inelastic x-ray scattering. These are topics of future interest.

\acknowledgments The research described in this publication was made
possible in part by Grant No. UP2-2436-LV-02 of the U.S. Civilian Research
and Development Foundation for the Independent States of the Former Soviet
Union (CRDF). J.K.F. also acknowledges support from the National Science
foundation under Grant No. DMR-0210717. T.P.D. would like to acknowledge
NSERC, PREA and the Alexander von Humboldt foundation for support of this
work.

\appendix

\section{Parquet contributions}

\begin{figure}[!htbf]
        \begin{center}
                \includegraphics[width=0.45\columnwidth]{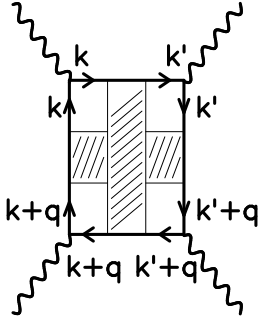}
        \end{center}
        \caption{\label{fig: parquet}
Feynman diagrams for a typical parquet-like renormalization. This resonant
diagram has a simultaneous horizontal and vertical renormalization by the
two-particle reducible charge vertex.  Note that such a renormalization is
only possible in the $A_{\textrm{1g}}$ sector.}
\end{figure}

In addition to the diagrams presented in Fig~\ref{fig:resonant_diagram},
there can also be parquet-like contributions with both vertical and
horizontal renormalizations. One type of these diagrams is shown in
Fig.~\ref{fig: parquet}. The corresponding expression has the form
\begin{align} \label{parq8}
        \frac{t^{*4}}{D^2} \frac1{N^3} \sum_{\substack{\bm {q k k'}\\\alpha\beta\alpha'\beta'}}
        \sin k_\alpha \sin (k_\beta + q_\beta) \sin k'_{\alpha'} \sin (k'_{\beta'} + q_{\beta'})
        \\
        \nonumber
        \times
        T^2\Gamma(1, 3)\Gamma(2, 4)
        G_{1}({\bm k'}) G_{2}({\bm {k'}})
 G_{2}({\bm {k'+q}}) G_{3}({\bm {k'+q}}) \\
\nonumber \times G_{2}({\bm {k+q}}) G_{4}({\bm {k+q}}) G_{4}({\bm {k}})
G_{1}({\bm {k}}) .
\end{align}
In the expression in Eq.~(\ref{parq8}), all the momentum dependence (in
the $D=\infty$ limit) is contained in the band energy [see
Eq.~(\ref{G_DMFT})] and, after expanding the products of the Green's
functions with the same momentum into partial fractions over
$\epsilon_{\bm k}$, the summations over momentum are of the form
\begin{align} \label{parq4}
        \frac{t^{*4}}{D^2} \frac1N \sum_{\bm {q}}
        \left[
        \frac1N \sum_{\bm {k}}\sum_{\alpha\beta}
        \frac{\sin k_\alpha \sin (k_\beta + q_\beta)}
        {(Z_1-\epsilon_{\bm k})(Z_2-\epsilon_{\bm {k+q}})}
        \right]
        \\
        \nonumber
        \times
        \left[
        \frac1N \sum_{\bm {k'}}\sum_{\alpha'\beta'}
        \frac{\sin k'_{\alpha'} \sin (k'_{\beta'} + q_{\beta'})}
        {(Z_3-\epsilon_{\bm {k'}})(Z_4-\epsilon_{\bm {k'+q}})}
        \right].
\end{align}

In the same way as was done for Eq.~(\ref{sum_k1}), we find the expression
in the bracket reduces to the following in the $D\to\infty$ limit
[$\zeta_i=\sgn(\Img Z_i)$]
\begin{align}\label{br1}
        i^2 \int_0^{\zeta_1\infty} d\lambda_1 \int_0^{\zeta_2\infty} d\lambda_2 \;
        e^{-i(\lambda_1 Z_1 + \lambda_2 Z_2)} \;
        e^{-t^2(\lambda_1^2+\lambda_2^2+2\lambda_1 \lambda_2 X_{\bm q})/4}
        \\
        \nonumber
        \times\left[
        \frac12 \sum_\alpha \cos q_\alpha +
        \frac{t^{*2}\lambda_1 \lambda_2}{4D} \sum_{\alpha\ne\beta} \sin q_\alpha \sin q_\beta
        \right],
\end{align}
where $X_{\bm q}=\lim_{D\rightarrow\infty}\frac1D\sum_{\alpha=1}^D \cos
q_\alpha$. The main contribution comes from the first term in
Eq.~(\ref{br1}) and there is a similar term in the second bracket of
Eq.~(\ref{parq4}). Replacing the square of a cosine by its average value
$\frac12$, we find that Eq.~(\ref{parq4}) reduces to
\begin{equation}
        \lim_{D\rightarrow\infty}\frac{t^{*4}}{8D} G_1 G_2 G_3 G_4 \to 0
\end{equation}
which vanishes as $D\rightarrow \infty$.  A similar procedure can be
performed for all other terms with a parquet-like renormalization.  Hence,
the parquet-like contributions are unimportant for resonant Raman
scattering in large dimensions.


\addcontentsline{toc}{section}{Bibliography}
\begin{thebibliography}{99}

\bibitem{Pap1}
P. Nyhus, S. L. Cooper, and Z. Fisk, Phys. Rev. B \textbf{51}, 15626
(1995).

\bibitem{Pap2}
P. Nyhus, S. L. Cooper, Z. Fisk, and J. Sarrao, Phys. Rev. B \textbf{52},
R14308 (1995); P. Nyhus, S. L. Cooper, Z. Fisk, and J. Sarrao, Phys. Rev.
B \textbf{55}, 12488 (1997).

\bibitem{Pap3}
H. L. Liu, S. Yoon, S. L. Cooper, G. Cao and J. E. Crow, Phys. Rev. B
\textbf{60}, R6980 (1999).

\bibitem{resonance}
K. B. Lyons, P. A. Fleury, L. T. Schneemeyer, and J. V. Waszczak, Phys.
Rev. Lett. \textbf{60}, 732 (1988); S. Sugai, S. Shamoto, and M. Sato,
Phys. Rev. B \textbf{38}, 6436 (1988); P. E. Sulewsky, P. A. Fleury, K. B.
Lyons, S.-W. Cheong, and Z. Fisk, Phys. Rev. B \textbf{41}, 225 (1990); R.
Liu, M. V. Klein, D. Salamon, S. L. Cooper, W. C. Lee, S.-W. Cheong, and
D. M. Ginsberg, J. Phys. Chem. Solids \textbf{54}, 1347 (1993).

\bibitem{Blumberg}
G. Blumberg, P. Abbamonte, M. V. Klein, W. C. Lee, D. M. Ginsberg, L. L.
Miller, and A. Zibold, Phys. Rev. B \textbf{53}, 11930 (1996).

\bibitem{irwin}
J. G. Naeini, X. K. Chen, J. C. Irwin, M. Okuya, T. Kimura, K. Kishio,
Phys. Rev. B {\bf 59}, 9642 (1999).

\bibitem{paper1}
J. K. Freericks and T. P. Devereaux, Condens. Matter Phys. \textbf{4}, 149
(2001); Phys. Rev. B \textbf{64}, 125110 (2001).

\bibitem{paper2}
J. K. Freericks, T. P. Devereaux, and R. Bulla, Acta Phys. Polon. B
\textbf{32}, 3219 (2001); Phys. Rev. B \textbf{64}, 233114 (2001); Acta
Phys. Polon. B \textbf{34}, 737 (2003); J. K. Freericks, T. P. Devereaux,
R. Bulla, and Th. Pruschke, Phys. Rev. B \textbf{67}, 155102 (2003).

\bibitem{paper3}
T. P. Devereaux, G. E. D. McCormack, and J. K. Freericks, Phys. Rev. Lett.
\textbf{90}, 067402 (2003); Phys. Rev. B \textbf{68}, 075105 (2003).

\bibitem{Chubukov}
A. V. Chubukov and D. M. Frenkel, Phys. Rev. B \textbf{52}, 9760 (1995);
Phys. Rev. Lett. \textbf{74}, 3057 (1995); D. K. Morr and A. V. Chubukov,
Phys Rev. B \textbf{56}, 9134   (1997).

\bibitem{Tohyama}
T. Tohyama, H. Onodera, K. Tsutsui, and S. Maekawa, Phys. Rev. Lett.
\textbf{89}, 257405 (2002).

\bibitem{LF}
P. A. Fleury and R. Loudon, Phys. Rev. \textbf{166}, 514 (1968).

\bibitem{RRP1}
R. R. P. Singh, Comments Condens. Matter Phys. \textbf{15}, 241 (1991).

\bibitem{Dagotto}
E. Dagotto and D. Poilblanc, Phys. Rev. B \textbf{42}, 7940 (1990); E.
Gagliano and S. Bacci, Phys. Rev. B \textbf{42}, 8772 (1990)

\bibitem{Sandvik}
A. W. Sandvik, S. Capponi, D. Poilblanc, and E. Dagotto, Phys. Rev. B
\textbf{57}, 8478 (1998).

\bibitem{Morr}
D. K. Morr, A. V. Chubukov, A. P. Kampf, and G. Blumberg, Phys. Rev. B
\textbf{54}, 3468 (1996).

\bibitem{RRP2}
R. R. P. Singh, P. A. Fleury, K. B. Lyons, and P. E. Sulewshi, Phys. Rev.
Lett. \textbf{62}, 2736 (1989).

\bibitem{CanaliGirvin}
C. M. Canali and S. M. Girvin, Phys. Rev. B \textbf{45}, 7127 (1992).

\bibitem{Kampf}
A. A. Katanin and A. P. Kampf, Phys. Rev. B \textbf{67}, 100404 (2003).

\bibitem{ll8979}
A. Shvaika, O. Vorobyov, J. K. Freericks, and T. P. Devereaux,
{\it to appear in} Phys. Rev. Lett.  (cond-mat/0311070).

\bibitem{FK}
L. M. Falicov and J. C. Kimball, Phys. Rev. Lett. \textbf{22}, 997 (1969).

\bibitem{Shastry}
B. S. Shastry and B. I. Shraiman, Phys. Rev. Lett. \textbf{65}, 1068
(1990); Int. J. Mod. Phys. B \textbf{5}, 365 (1991).

\bibitem{BM}
U. Brandt and C. Mielsch, Z. Phys. B \textbf{75}, 365 (1989); \textbf{79},
295 (1990).

\bibitem{SFM}
A. M. Shvaika, Physica C \textbf{341-348}, 177 (2000); J. K. Freericks and
P. Miller, Phys. Rev. B \textbf{62}, 10022 (2000); A. M. Shvaika, J. Phys.
Studies \textbf{5}, 349 (2001).

\bibitem{shvaika_conf1} 
A. Shvaika, O. Vorobyov, J. K. Freericks, and T. P. Devereaux,
{\it submitted to} J. Phys. Chem. Solids (cond-mat/0407120).

\bibitem{shvaika_conf2}
A. Shvaika, O. Vorobyov, J. K. Freericks, and T. P. Devereaux,
{\it to appear in} Physica B (cond-mat/0406305).

\bibitem{MartinFalicov} R. M. Martin and L. M. Falicov, in \emph{Light Scattering in Solids}, edited by M.
Cardona (Springer-Verlag, New York, 1975).

\bibitem{prl_mit} D. O. Demchencko, A. V. Joura, and  J. K. Freericks, 
Phys. Rev. Lett. {\bf 92}, 216401 (2004).

\bibitem{prb_thermal}
 J. K. Freericks, D. Demchenko, A. Joura, and V. Zlati\'c, Phys. Rev. B 
{\bf 68}, 195120 (2003).

\end{thebibliography}
\end{document}